\documentclass[usenatbib,preprint]{mn2e}

\usepackage{graphicx}
\usepackage {natbib,aas_macros}
\usepackage {bm}
\usepackage {amsmath,amssymb}
\usepackage {setspace}

\newcommand{\cm}{~{\rm g~cm}^{-3} } 

\title[Effects of radiative transfer on the structure of self-gravitating disks]{Effects of radiative transfer on the structure of self-gravitating disks, their fragmentation and evolution of the fragments}

\author[Tsukamoto et al]{Y. Tsukamoto$^{1}$,S. Z. Takahashi$^{1}$, 
M. N. Machida$^{2}$, and  S. Inutsuka$^{1}$ \\
$^1$Department of Physics, Nagoya University, Furo-cho, Chikusa-ku, Nagoya, Aichi, Japan  \\
$^2$Department of Earth and Planetary Sciences, Kyushu University, 
6-10-1 Hakozaki, Higashi-ku, Fukuoka, Fukuoka, Japan \\
}

\begin{document}
\maketitle

\begin{abstract}
We investigate structure of self-gravitating disks, their fragmentation and
evolution of the fragments (the clumps) using both analytic approach and three-dimensional radiation
hydrodynamics simulations starting from molecular cores.
The simulations show that non-local radiative transfer determines disk temperature.
We find the disk structure is well described by an analytical model of
quasi-steady self-gravitating disk with radial radiative transfer.
Because the radiative process is not local and radiation from the interstellar medium
cannot be ignored, the local radiative cooling would not be balanced with the
viscous heating in a massive disk around a low mass star.
In our simulations, there are cases in which the disk does not fragment even though it satisfies the 
fragmentation criterion based on disk cooling time ($Q \sim 1$ and $\Omega t_{\rm cool}\sim 1$). 
This indicates that at least the criterion is not sufficient condition for fragmentation.
We determine the parameter range for the host cloud core in which disk fragmentation occurs.
In addition, we show that the temperature evolution of the center of the clump 
is close to that of typical first cores and 
the minimum initial mass of clumps to be about a few Jupiter mass. 
\end{abstract}


\section{Introduction}
\label{intro}
Stars form in gravitationally collapsing molecular cloud cores.
Since molecular cloud cores have angular 
momentum \citep{1993ApJ...406..528G,2002ApJ...572..238C}, 
a circumstellar disk forms around the protostar.
According to recent theoretical studies of the gravitational collapse of 
molecular cloud cores, the protostar is surrounded by a circumstellar 
disk early in its evolutionary phase in the case without magnetic field,
\citep{1998ApJ...508L..95B,2010ApJ...724.1006M,2011MNRAS.416..591T} and with magnetic field 
\citep{2010ApJ...718L..58I,2011ApJ...729...42M}.

As \citet{2010ApJ...718L..58I} pointed out, 
a circumstellar disk can be gravitationally unstable during its early evolution.
When the protostar forms,  its mass is only  $10^{-2}$ to $10^{-3}~M_{\odot}$. 
In contrast, the first-core that is the precursor to the 
circumstellar disk \citep{2006ApJ...645..381S,2010ApJ...724.1006M} 
has an initial mass of $\sim 0.1~M_{\odot}$. Therefore, in the formation epoch of the protostar,  
the disk has a mass that is significantly greater than that of the protostar.
Observations also suggest that 
massive disks can form in the main accretion phase \citep{2009ApJ...707..103E}.

In such massive disks, gravitational instability (GI) can develop. 
The Toomre's  $Q$ value,
\begin{eqnarray}
Q=\frac{\kappa_{\rm ep} c_s}{\pi G \Sigma},
\end{eqnarray}
is a well-known indicator for GI \citep{1964ApJ...139.1217T}. 
When $Q\lesssim 1.5$, the disk is gravitationally unstable
against non-axisymmetric perturbations and develops 
spiral arms \citep{1994ApJ...436..335L}. 
The spiral arms readjust the mass and angular 
momentum of the disk, promoting mass accretion onto the protostar.
They also heat the disk gas.
By readjusting the surface density and the disk heating, 
the $Q$ value increases, and the disk is stabilized.
This self-regulative nature is an important aspect of GI.
To maintain GI, additional physical processes 
such as radiative cooling or mass accretion from the envelope are necessary.
With these effects, the disk can also fragment and
gravitationally bound gaseous objects (which we call  ``clumps") form.

Disk fragmentation is a candidate mechanism for the formation 
of wide-orbit planets
\citep{2010ApJ...714L.133V,2011ApJ...729...42M,2013ApJ...768..131V}.
Wide-orbit planets are planets separated from the central star by more than 
10 AU \citep{2008Sci...322.1348M,2009ApJ...707L.123T,
2009A&A...493L..21L,2010Natur.468.1080M,2010ApJ...719..497L}.
Furthermore, it has been suggested that disk fragmentation 
can explain the formation of brown dwarfs 
\citep{2009MNRAS.392..413S,2011MNRAS.413.1787S} and multiple stellar systems
\citep{2008ApJ...677..327M,2010ApJ...708.1585K}.

Effects of radiative cooling on GI and disk fragmentation 
has been investigated in \citet{2001ApJ...553..174G} with
two dimensional local shearing box simulations. 
To model radiative cooling, he employed a simplified cooling law (see right hand side of equation (\ref{sec2_cooling0})).
In his simulations, the disk is initially unstable against GI ($Q =1$) and
GI immediately develops and heats the disk. When radiative cooling is not so strong,
it balances the heating by GI, and the disk settles into a quasi-steady state. 
This quasi-steady state also realizes in three dimensional global disk simulations \citep{2004MNRAS.351..630L}.
On the other hand, when radiative cooling is sufficiently strong, 
such a quasi-steady state can not be realized and the disk fragments.
In the simulations of \citet{2001ApJ...553..174G}, disk fragmentation occurred when disk cooling time $t_{\rm cool}$
is comparable to the orbital timescale, $t_{\rm cool} \sim t_{\rm orbit}$ (or $\Omega t_{\rm cool}\sim O(1)$).
The fragmentation condition are also confirmed by three dimensional 
global disk simulations \citep{2003MNRAS.339.1025R}
but the simplified cooling laws are also used.

Since \citet{2001ApJ...553..174G} and \citet{2004MNRAS.351..630L} employed the simplified cooling law,
they implicitly assumed that radiation just acts as the cooling process
to decrease the local gas energy (we call this 
the assumption of local radiative cooling).
However, this assumption is not necessarily true. 
For example, irradiation from the protostar and that
from the inner disk region can heat the disk.
Thus, in reality, incoming radiation flux, in addition to outgoing radiation flux and
local GI heating, should be considered.
Furthermore, the interstellar medium has a typical 
temperature of about 10 K and it is almost impossible to cool the disk gas 
below 10 K because of radiation flux from the ambient interstellar medium.
Therefore, it is not clear that, in realistic situations, 
the local balance between radiative cooling and viscous heating
is achieved or not as in \citet{2001ApJ...553..174G}.
Whether the balance is achieved is very important because
the structure of quasi-steady self gravitating disk can be determined
by the energy balance.

Another important issue is  applicability of the fragmentation 
criterion found by \citet{2001ApJ...553..174G},
$Q \sim 1$ and $\Omega t_{\rm cool} \sim 1$, in realistic situations.
Gammie pointed out that, in realistic systems, 
fragmentation realizes when the external irradiation
quickly diminishes and when the gas quickly cools.
Thus, he regarded the fragmentation 
criterion can be applicable in very limiting cases.

On the other hand, \citet{2003MNRAS.339.1025R} interpreted this criterion as
``almost isothermal conditions are necessary for fragmentation". 
When the cooling timescale of the disk is small, the gas evolves isothermally during GI growing and 
pressure repulsion becomes weak compared to the adiabatic evolution case (or inefficient cooling case).
\citet{2003MNRAS.339.1025R} suggested that such  almost isothermal evolution in non-linear evolution phase of GI
is necessary for fragmentation. According to this interpretation, how the disk becomes $Q\sim 1$ or the energy
balance within it is not important because whether fragmentation realize or not depends on the thermal behavior of
the gas in the non-linear evolution of GI. The criterion
seems to be regarded as a necessary condition in this interpretation.

In the previous works using analytic approach, however, the criterion is used as if
a sufficient condition for fragmentation \citep{2005ApJ...621L..69R,2011MNRAS.417.1928F,2011ApJ...740....1K,2013MNRAS.432.3168F}.
As just described above, the interpretation of the criterion is ambiguous and its applicability 
in the realistic situation is still not clear.

Another mechanism that makes a disk gravitationally unstable is mass loading from 
the envelope \citep[e.g.,][]{2003ApJ...595..913M,2010ApJ...718L..58I,2010ApJ...708.1585K,
2010ApJ...714L.133V,2011ApJ...729...42M,
2011MNRAS.416..591T,2011MNRAS.417.2036B,2011ApJ...730...32S,2013MNRAS.436.1667T,2013MNRAS.428.1321T}. 
This mechanism is efficient in the early phase of disk 
evolution during which the protostar and disk are embedded in a massive envelope. 
With mass accretion from the envelope, the
$Q$ value of the disk decreases due to the increase of 
the surface density. When the mass accretion onto the disk is sufficiently high,
the disk eventually fragments, and planetary mass clumps form. 
In this process, whether fragmentation occurs depends
on envelope (or cloud core) parameters such as thermal energy or rotational energy.
The parameter range in which
fragmentation occurs has already been investigated in our previous 
studies \citep{2011MNRAS.416..591T,2013MNRAS.428.1321T}; however, in those studies,
the effects of radiative transfer were approximated.
Radiative transfer could 
also play important roles during the early phase of disk
evolution; therefore, a parameter survey of disk 
fragmentation with radiation hydrodynamics simulation is necessary. 

Initial properties and the evolution of fragments (or clumps) are other important issues. 
The ultimate fate of clumps depends on their orbital and internal evolutions.
For example,
if the radial migration timescale is very short,
clumps quickly accrete onto the protostar and disappear.
If the radial migration timescale is sufficiently long and the clumps survive in the disk
maintaining its mass in the range of planetary masses ($\lesssim 10 M_{\rm Jupiter}$), the clumps evolve into
the wide orbit planets \citep{2010ApJ...714L.133V,2011ApJ...729...42M}.
If the mass of the clumps quickly increases, disk fragmentation
can be regarded as the formation process for brown dwarfs and
binary or multiple stellar systems \citep{2009MNRAS.392..413S}. 

In spite of its importance, only a few studies have focused on the orbital and 
internal evolution of clumps in circumstellar disks.
\citet{2011MNRAS.416.1971B} investigated the orbital evolution of 
massive planets in a gravitationally unstable disk under the 
assumption of local radiative cooling.
They showed that massive planets rapidly migrate inward 
in a  type I migration 
timescale \citep{2002ApJ...565.1257T}. 
By adopting an analytical approach or 
three-dimensional smoothed particle hydrodynamics (SPH)
simulation, \citet{2010MNRAS.408.2381N} and \citet{2012MNRAS.427.1725G}
investigated the internal evolution 
and the collapse timescale of an isolated clump. 
The gravitational collapse  (the second collapse) 
occurs in the clump when the 
central temperature reaches $\sim 2000$ K,
at which the dissociation of molecular
hydrogen begins, and the 
gas pressure can no longer support the clump against its self gravity. 
These authors suggested that the timescale for the second collapse 
after clump formation 
is in the range of several thousand years to several 
$10^4$ years. However, they ignored further 
mass accretion from the disk onto the clumps.

\citet{2013MNRAS.436.1667T} investigated the orbital and internal 
evolution of clumps permitting realistic gas accretion
from the disk onto clumps with three-dimensional radiation hydrodynamics simulation.
They showed that although most of the clumps rapidly migrate 
inward and finally fall
onto the  protostar, a few clumps can remain in the disk.
They also showed that clumps are convectionally stable ($\gamma>\gamma_{\rm eff}$).
The central density and temperature of a surviving 
clump rapidly increase, and the clump undergoes a second collapse 
within $1000$ -- $2000$ years after its formation.   
However, in this works, only one simulation was performed. 
The evolution process of clumps may change under different
initial conditions. For example, the central entropy of the clump
would become small when  fragmentation occurs in the outer relatively 
cold region of the disk. Since the central entropy of the clump determines its initial
mass and radius of the clump, fragmentation in different disk environments
changes the nature of the clumps.
It is also expected that the mass accretion from the disk onto clumps becomes small when fragmentation occurs in
the outer region where the disk surface density is small at $Q=1$.

In this study, we investigate the structure of self-gravitating disks,
fragmentation of the disk, and evolution of clumps using
both an analytic approach and radiation hydrodynamics simulations 
starting from molecular cloud cores.
This paper is organized as follows. In \S 2, we analytically derive 
the structure of the self-gravitating steady disk using various energy balance relations. 
It would be insightful to understand the general structure 
of self-gravitating disks.
The numerical method and initial conditions for the simulations are
described in \S 3.
Results and discussions are divided into two parts (\S \ref{self_grav_disks} and \S \ref{fragment_section}). 
In \S \ref{self_grav_disks}, we
mainly investigate the structure of a self-gravitating 
disk that formed in the radiation hydrodynamics simulations. 
The simulation results for disk evolution
are presented in \S \ref{results_disk}.
In this section, we show that radiation can 
transfer energy within the disk and can heat 
the outer region of the disk.
Therefore, the assumption of local radiative
cooling is not valid in the disk.
In \S \ref{self_grav_disk_discussion}, we discuss the consequences of the results
obtained from our numerical simulations
and their implications in interpreting previous work.
In \S \ref{fragment_section}, we investigate 
disk fragmentation and the evolution 
of clumps. In \S \ref{fragment_simulation_results},  simulation results are shown.
The parameter range of the molecular cloud core  in which fragmentation occurs is 
determined, and the orbital and internal evolutions of clumps 
are investigated.
In \S \ref{fragment_simulation_discussion}, 
we discuss the results reported in \S \ref{fragment_simulation_results}
and derive the minimum initial mass of clumps.
Finally, we summarize our results and discuss future perspectives in \S \ref{summary_future}.

\section{Structures of self-gravitating steady disks with various energy balance equations}
\label{theory_disk}
The analytical description of the radial structure of a self-gravitating steady disk is useful for interpreting
the simulation results shown in this study and in previous studies.
In this section, we derive radial profiles for physical quantities in
a self-gravitating circumstellar disk using various energy balance equations.
We  assume that
the disk (1) is steady ($\dot{M}={\rm const.}$), (2) can be described by the
viscous $\alpha$ accretion disk model \citep{1973A&A....24..337S} and
(3) is in the quasi steady-state against gravitational instability ($Q\sim 1$).

A steady viscous accretion disk should satisfy the following equations,
\begin{equation}
\label{sec2_dynamics}
\begin{split}
\dot{M} = -2 \pi r v_r \Sigma = {\rm const.}~ (\propto r^0), \\
\left| \frac{d \ln \Omega}{d \ln R} \right| \alpha \frac{c_s^2}{\Omega} \Sigma = \frac{1}{2 \pi}\dot{M},
\end{split}
\end{equation}
where, $\alpha= \nu \frac{\Omega}{c_s^2}$, and $\nu$ is the kinematic viscosity.
The $Q$ for a self-gravitating disk becomes
\begin{equation}
\label{sec2_qvalue}
Q=\frac{\kappa_{ep} c_s}{\pi G \Sigma}\sim 1 ~( \propto r^0).
\end{equation}
Here and in the following, we assume the epicycle frequency
 $\kappa_{\rm ep}$ scales the same as the angular velocity  $\Omega$.
We assume that physical quantities obey the single power law in radius,
$\Sigma \propto r^{n_\Sigma}$,$T \propto r^{n_T}$, $\Omega \propto r^{n_\Omega}$,
and $\alpha \propto r^{n_\alpha}$.
Then, (\ref{sec2_dynamics}) and (\ref{sec2_qvalue}) lead to
\begin{equation}
\label{eq_steady1}
\begin{split}
n_\Sigma=\frac{1}{2} \left( n_T+2 n_\Omega \right), \\
n_\alpha = -\frac{3}{2} n_T.
\end{split}
\end{equation}
From these equations, we can determine the 
profile of a steady self-gravitating disk 
by specifying a rotation profile and an energy balance equation.

Equations (\ref{eq_steady1}) shows that
only a globally isothermal disk can achieve a
globally constant $\alpha$. In other words, 
if the disk has a radial temperature profile, then
it also inevitably has a radial profile for $\alpha$.
Therefore, all  self-gravitating disk models
that have a radially constant 
$\alpha$ and a radially dependent temperature profile violate
the assumption of steady state.

\subsection{Disk structure with local cooling law}
First, we investigate the steady-state structure of a disk with
local cooling law \citep{2001ApJ...553..174G}, 
which is often used in global disk simulations
in the context of gravitational instabilities
or disk fragmentation
\citep[e.g.,][]{2005MNRAS.364L..56R,2005MNRAS.358.1489L,2011MNRAS.416L..65P,2012MNRAS.427.2022M}.

At the steady state with local cooling law, the energy balance is given by
relation,
\begin{equation}
\label{sec2_cooling0}
 \left| \frac{d \ln \Omega}{d \ln R} \right|^{2} \alpha \frac{c_s^2}{\Omega}  \Sigma \Omega^2  = \frac{\Sigma c_s^2}{\gamma(\gamma-1)t_{\rm cool}}.
\end{equation}
This can be rewritten as 
\begin{equation}
\label{sec2_cooling1}
\alpha = \left| \frac{d \ln \Omega}{d \ln R} \right|^{-2} \frac{1}{\gamma(\gamma-1)\Omega t_{\rm cool}} \propto r^0,
\end{equation}
where, we adopted $\Omega t_{\rm cool} \equiv \beta={\rm const.} \propto r^0$ 
as in most previous studies and equation (\ref{sec2_cooling1})
gives a disk with constant $\alpha$.
Note that there is no reason to expect that $t_{\rm cool}$ is constant.

Combining (\ref{sec2_dynamics}), (\ref{sec2_qvalue}), and (\ref{sec2_cooling1}), we obtain
\begin{equation}
\label{sec2_gammie}
\alpha \propto r^0,~\Sigma \propto r^{-\frac{3}{2}},~T \propto r^0.
\end{equation}
Here, we assumed Keplerian disk,  $n_\Omega = -\frac{3}{2}$. 
This cooling law gives a globally isothermal disk and a
surface density of $\Sigma \propto r^{-\frac{3}{2}}$.
Indeed, Fig. 1 in \citet{2011MNRAS.416.1971B} shows that
the disk is almost globally isothermal 
and that the surface density 
has a profile of $\Sigma \propto r^{-\frac{3}{2}}$ in the quasi-steady state of their simulations.
Their results are consistent with our estimate.


Because the disk is isothermal, we can easily calculate the 
disk temperature in the quasi-steady state (we call this ``the equilibrium temperature") 
using physical quantities at a radius.
For example, we can calculate the temperature of the disk used in \citet{2005MNRAS.364L..56R} and \citet{2012MNRAS.427.2022M}.
Since dimensionless disk parameters were used in those studies, we need to convert them into dimensional
parameters to derive dimensional disk quantities.
If we regard the disk model used in those studies as
a disk with $M_{\rm star}=1 M_{\odot},~M_{\rm disk}=0.1 M_{\odot}$, the disk cutoff 
radius as $r_{\rm in}=0.25 $ AU and $r_{\rm out}=25$ AU as suggested by \citet{2012MNRAS.427.2022M}, then
the equilibrium temperature $T_{\rm eq}$ can be calculated by substituting the value of the physical quantities into
\begin{equation}
T_{\rm eq}=\left( \frac{\mu m_{\rm H}}{\gamma k_B}\right )\left(\frac{Q \pi G \Sigma}{\Omega}\right)^{2}=4.7 K.
\end{equation}
Here we have used $Q=1$, and  $\mu=2.38$ is the mean molecular weight, $m_{\rm H}$ is the mass of a 
hydrogen atom, $k_B$ is the Boltzmann constant, and $\gamma=5/3$ is the ratio of heat capacities.
This equilibrium temperature is too small for a disk temperature around a low-mass star. For example,
the temperature of disk at $10$ AU is about $100$ K if the luminosity of the central star is $1L_\odot$. Note also that the disk 
is very compact and condensed ($0.1 M_{\odot}$ within $r_{\rm out}=25$ AU).
This disk has a typical Jeans mass of
\begin{equation}
M_{\rm Jeans}\sim10^{-5} M_{\odot},
\end{equation}
(see, top panel of Fig. \ref{resolution2}).
Therefore, the mass of the fragments formed in their simulations would be about
$0.01 M_{\rm Jupiter}$.
Note that, because the Jeans mass is very small, 
the resolution requirement for such a disk is very severe 
and $N\gtrsim 10^7$ particles are required for an SPH simulation to resolve the Jeans mass and 
satisfy the resolution requirement suggested by \citet{1997MNRAS.288.1060B}
(see, \S \ref{sec5_res} for more discussion).

\subsection{Disk structure with the local energy balance equation}
Next, we investigate the disk structure when
local viscous heating balances local radiative cooling.
This assumption is also often used in the context of disk fragmentation
\citep[e.g.,][]{2005ApJ...621L..69R,2011ApJ...740....1K}.
The local energy balance equation between radiative cooling and 
viscous heating is given by
\begin{equation}
\label{sec2_cooling2}
\begin{split}
 \left| \frac{d \ln \Omega}{d \ln R} \right|^{2} \alpha \frac{c_s^2}{\Omega}  \Sigma \Omega^2 = \frac{32 \sigma T^4}{3 \tau},
\end{split}
\end{equation}
where $\tau$ is the optical depth and we assume $\tau=\frac{1}{2} \kappa \Sigma$.
Here we assume that the disk is optically thick.
When the gas temperature below $100$ K, the opacity  $\kappa$ is well described by 
\begin{equation}
\label{sec2_opacity}
\kappa=\kappa_0 T^2,
\end{equation}
where $\kappa_0$ is a constant \citep{2003A&A...410..611S}.
As we will see below, the disks formed in our simulation 
satisfy the condition, $T< 100$ K for $r \gtrsim 10$ AU. 

Using (\ref{sec2_dynamics}), (\ref{sec2_qvalue}), and (\ref{sec2_cooling2}), 
we have 
\begin{equation}
\label{sec2_local_cooling}
\begin{split}
\Sigma \propto r^{-3},~T \propto r^{-3},~\alpha \propto r^{\frac{9}{2}}.
\end{split}
\end{equation}
where we again assume the disk is Keplerian, $n_\Omega=-\frac{3}{2}$.
With this energy balance equation,
physical quantities have very steep radial profiles.
Especially, the radial dependence of the temperature ($T \propto r^{-3}$)
is significantly steeper than the profile expected from the passively irradiated disk 
model, $T\propto r^{-\frac{3}{7}}$ \citep[see,][]{1970PThPh..44.1580K,1997ApJ...490..368C}. 
Whether such steep profiles actually realize in realistic situations is unclear, and we must
investigate the disk profile with three-dimensional radiation hydrodynamics simulations.

The exponents in the power laws for the disk profile are determined 
by (\ref{sec2_dynamics}), (\ref{sec2_qvalue}), 
the energy balance equation (\ref{sec2_cooling2}),
and the rotation profile $\Omega$. Thus, there
is no degree of freedom available to change them.
Therefore, it is impossible to realize a radially constant 
$\alpha$ in the Keplerian disk with local
energy balance equation.


\subsection{Disk structure with a given temperature profile}
As we will see below, the radiation flux from the inner region can heat
the outer region of the disk. Furthermore, in a realistic disk 
around a protostar, stellar irradiation may have an important effect on the disk temperature.
Therefore, it is useful to investigate the steady-state structure
of the disk with a given temperature profile.

In a disk whose temperature is passively determined,
the temperature profile can be written as
\begin{equation}
T \propto r^{n_{\rm given}},
\end{equation}
instead of using an energy balance equation, such as equation (\ref{sec2_cooling1}) or (\ref{sec2_cooling2}).
The value of $n_{\rm given}$ depends on the disk model. For example, 
in the irradiated disk model, $n_{\rm given}=-\frac{3}{7}$ \citep[see,][]{1970PThPh..44.1580K,1997ApJ...490..368C}.
Using (\ref{sec2_dynamics}), (\ref{sec2_qvalue}), $n_{\rm given}=-\frac{3}{7}$, and $n_\Omega=-\frac{3}{2}$,
we can derive the structure of the irradiated steady self-gravitating disk, 
\begin{equation}
\label{given_temp}
\begin{split}
\Sigma \propto r^{-12/7},~T \propto r^{-\frac{3}{7}},~\alpha \propto r^{9/14}.
\end{split}
\end{equation}
This structure is expected when stellar irradiation directly reaches the disk photosphere. 
If the radiation is extinct before it reaches  the disk photosphere, 
the temperature profile becomes steeper than that given in (\ref{given_temp}).
With these results in mind, we investigate the disk structure observed in three-dimensional radiation hydrodynamical
simulations.

\section{Numerical Method and Initial Conditions}
\label{method}
In the simulations performed for this study, 
we solved the radiation hydrodynamics equations with the flux-limited diffusion (FLD) approximation 
SPH schemes according to
\citet{2004MNRAS.353.1078W} and \citet{2005MNRAS.364.1367W}. 
We adopted the tabulated equation of state  used in
\citet{2013ApJ...763....6T}, in which the internal degrees of freedom 
and chemical reactions of seven species ${\rm H_2,~H,~H^+,~He,~He^+,He^{++}, e^-}$ are included.
The hydrogen and helium mass fractions were assumed to be $X=0.7$ and $Y=0.28$, respectively.
We used the dust opacity table provided by \citet{2003A&A...410..611S} 
and the gas opacity table from \citet{2005ApJ...623..585F}.

When the gas density exceeds the threshold
density, $\rho_{{\rm thr}} ~(5 \times 10^{-8} ~{\rm g\,cm^{-3}})$,
a sink particle was introduced.
Around this density, the gas temperature reaches the dissociation 
temperature of molecular hydrogen ($T\sim 1500$\,K) and the second
collapse begins in the first-core or in the clumps. 
Therefore, we can follow the thermal evolution 
just prior to the second collapse.
The sink radius was set to $r_{{\rm sink}}=2$ AU.

We adopted an isothermal, uniform and rigidly rotating
molecular cloud core as the initial condition. 
Starting the simulations from  molecular cloud cores has several merits for investigating the disk evolution.
First, we can investigate the formation and evolution of the disk in a self-consistent manner.
Second, the boundary conditions for radiative transfer can be placed far from the disk photosphere.
As \citet{2011MNRAS.414..913R}  pointed out, the treatment of the boundary conditions significantly affects 
disk fragmentation in radiation hydrodynamics simulations.
Thus, an appropriate treatment for boundary conditions is crucial to investigate both the temperature structure 
of the disk and disk fragmentation.
For the boundary conditions, in this work, we fixed the gas temperature to be 10 K if 
the gas density is less than $10^{-18} \cm$. Thus, the boundary
was far from the disk photosphere and did not affect the temperature structure of the disk.

The initial mass and temperature of the cores
were 1 $M_\odot$ and 10 K; therefore, the free parameters of the core were the radius $R_0$
and the angular velocity $\Omega_0$. The parameters of the initial cloud cores are listed in Table 1.
In this table, the values of the initial condition used \citet{2013MNRAS.436.1667T} (model TMI13) 
is also shown to allow us to investigate the boundary between fragmentation and no-fragmentation models.
In the table, $\alpha_{\rm th}$ and $\beta_{\rm rot}$ are the 
ratios of thermal to gravitational energy ($\alpha_{\rm th}$ $\equiv E_{\rm t}/|E_{\rm g}|$)
and rotational to gravitational energy ($\beta_{\rm rot}$ $\equiv E_{\rm r}/|E_{\rm g}|$), respectively.
The values for cloud core parameters adopted in this study are consistent with results obtained by
recent three-dimensional magneto-hydrodynamics (MHD) simulations
that investigated the evolution of the molecular cloud and involved 
core formation \citep{2012ApJ...759...35I}.
The initial cores were modeled with about 520,000 SPH particles.

\begin{table*}
\label{initial_conditions}

\begin{center}
\caption{Model parameters and the simulation results. $R,~\Omega_{\rm init}$ and  $\rho_{\rm init}$ are
the radius, angular velocity and density of the initial cloud core, respectively.}		
\begin{tabular}{ccccccc}
\hline\hline
 Model & {\it R} (AU) & $\Omega_{\rm init} ({\rm sec}^{-1})$ & $\rho_{\rm init} ~({\rm g~cm^{-3}})$ & $\alpha_{\rm th}$ & $\beta_{\rm rot}$ & Fragmentation \\
\hline
 1  & $5.8 \times 10^3$ & $7.61\times 10^{-14}$ & $6.91\times 10^{-19}$ & 0.58  & 0.01  & No  \\
 2  & $5.8 \times 10^3$ & $1.07\times 10^{-13}$ & $6.91\times 10^{-19}$ & 0.58  & 0.02  & Yes  \\
 3  & $4.9 \times 10^3$ & $8.37\times 10^{-14}$ & $1.19\times 10^{-18}$  & 0.48 & 0.007 & No  \\
 4  & $4.9 \times 10^3$ & $1.00\times 10^{-13}$ & $1.19\times 10^{-18}$  & 0.48 & 0.01  & Yes  \\
 5  & $3.9 \times 10^3$ & $1.17\times 10^{-13}$ & $2.33\times 10^{-18}$  & 0.38 & 0.007 & No  \\
 TMI13  &  $3.9 \times 10^3$ & $1.40\times 10^{-13}$ & $2.33\times 10^{-18}$  & 0.38   & 0.01  & Yes  \\

\hline
\end{tabular}
\end{center}
\footnotesize
\end{table*}

\section{Quasi-steady structures of self gravitating disks with radiative transfer}
\label{self_grav_disks}
\subsection{Simulation results}
\label{results_disk}

\subsubsection{Overview of disk evolution}
We calculated the early evolution of the disk until 
about 10,000 years after the formation
of protostar.
Figure \ref{faceon_sigma} shows the evolution of surface density at the center
of the cloud core for model 1 in Table 1. In Figs.~\ref{faceon_sigma}{\it a} to \ref{faceon_sigma}{\it e}, we can clearly see the bimodal
structure, {\it i.e.,} the central pressure supported 
core (the first-core) and the disk around it. The figure indicates that, {\it before} the second collapse, the
disk forms around the central first-core, and spiral arms develop.
These features are commonly seen in the unmagnetized or weakly magnetized molecular cloud cores
\citep{1998ApJ...508L..95B,2011MNRAS.416..591T,2011MNRAS.417.2036B,2010ApJ...724.1006M,2011ApJ...729...42M,
2013MNRAS.436.1667T,2014MNRAS.437...77B}.
The central temperature of the first-core exceeds $T\sim 1500$ K and 
the second core (or protostar) 
forms between Figs.~\ref{faceon_sigma}{\it e} and \ref{faceon_sigma}{\it f}.
The white dot at the center of Figs.~\ref{faceon_sigma}{\it f} 
to \ref{faceon_sigma}{\it i} corresponds to the location of the protostar.  
The disk radius gradually increases 
by mass accretion from the envelope, and angular momentum 
is redistributed by the spiral arms.

Figure \ref{faceon_Tgas} shows gas temperatures at 
the center of the cloud at the same epochs as
in Fig. \ref{faceon_sigma}.
Comparing with Fig.~\ref{faceon_sigma}, Fig.~\ref{faceon_Tgas} shows that
the structure of the gas temperature is more axisymmetric than that of the surface density.
Figures \ref{faceon_sigma} and \ref{faceon_Tgas} indicate that 
there is only a very weak correlation between surface density and temperature. 
If we assume that the heat source for the disk is viscous heating due to the 
gravitational instability, it is difficult to explain the axisymmetric temperature
distribution because gravitational energy 
mainly converts to thermal energy around the spiral arms and the heating should be
localized around the spiral arms, and thus, the temperature 
structure should trace the surface density structure. 
Although we can see weak heating by the 
spiral arms (e.g., in Fig. \ref{faceon_Tgas}{\it g}), the entire temperature structure is almost axisymmetric.
Therefore, there must be other heating mechanisms that causes the axisymmetric temperature structure.
As we will see below, the radiation flux within the disk
determines the axisymmetric component of the gas temperature.

\subsubsection{Vertical disk structures}
Figure \ref{rz_map} shows the density contour map on the $y=0$ plane 
at the epoch in Fig.~\ref{faceon_sigma}{\it i}. 
The red lines show the scale height of disk, 
$H(r)=\frac{c_{s, {\rm mid}}(r)}{\Omega(r)}$, where
$c_{s, {\rm mid}}$ is the sound velocity on the 
mid-plane, and $\Omega$ is the angular velocity at $r$.
The cyan lines show the height of the vertical photosphere, $z_{\rm photo}$ defined as
 $\tau_z=\int^{\infty}_{z_{\rm photo}} \kappa \rho dz=1$.
The green lines and arrows show the temperature contour and direction of the radiation flux, respectively.  
The green lines show that, in the outer region, ($r \gtrsim 20$ AU), the disk is almost vertically isothermal. 
This feature also occurs in two-dimensional radiation hydrodynamics calculations \citep{1993ApJ...411..274Y}.
The green arrows around the mid-plane are almost parallel to the x-axis. 
This indicates that the radial component of the  radiation 
flux dominates the vertical component in the outer disk region.
Therefore, the radial component of radiative transfer 
play an important role in determining the temperature structure of the disk. 
We can see weak heating due to the spiral arms at $(x,z)=(40 ~{\rm AU},0 ~{\rm AU})$ where the green arrows spread out. 
Although such  spiral heating occasionally occurs, it is transient and does not significantly contribute to 
the temperature in the outer disk region.
Actually, as we will see below, the theoretical disk model in which local heating due to GI
balances radiative cooling cannot reproduce the radial profiles of this disk (see thick dashed lines in 
Fig. \ref{radial_profile}).

The vertical density structure is well described by the vertically isothermal 
thin disk profile, $\rho(z)=\rho_0 \exp(-z^2/2 H(r)^2)$,  although the disk is 
slightly compressed by the ram pressure of mass accretion from the envelope. 
The aspect ratio of the disk is about 0.2 at $r = 50$ AU.
In the inner region, $r \lesssim 30 $ AU, the height of the disk photosphere 
is greater than the disk scale height.
This structure is  different from passively irradiated 
optically thick disk \citep{1970PThPh..44.1580K}.
This is one reason why our disks have a steeper profile $T(r)\propto r^{-1.1} $ (see
left middle panel in Fig. \ref{radial_profile}) 
than the passively irradiated disk model $T(r) \propto r^{-\frac{3}{7}}$.

\subsubsection{Radial disk structure}
\begin{figure}
\includegraphics[width=90mm]{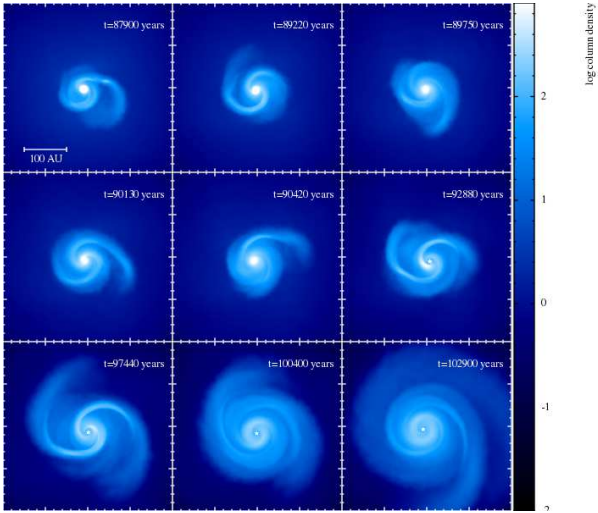}
\caption{
Time sequence for  surface density at the cloud center, viewed face-on for model 1.
Elapsed time after the cloud core begins to collapse is shown in each panel. 
}
\label{faceon_sigma}
\end{figure}

Figure \ref{radial_profile} shows the azimuthally averaged radial profiles of the disk.
Thin solid, dashed, dotted and dashed-dotted lines are for profiles at the same epochs as in
Figs.~\ref{faceon_sigma}{\it f}, {\it g}, {\it h} and {\it i}, respectively.
In the figure, top left panel shows the angular velocity $\Omega$. Because the disk is massive, the rotation
profile is not Keplerian ($\Omega \propto r^{-1.5}$; thick dotted line). 
We fitted $\Omega$ to $\Omega \propto r^{-1.1}$ 
(thick dotted-dashed line)
to obtain the power law of the disk with (\ref{sec2_dynamics}) and (\ref{sec2_qvalue}).

The middle left panel shows $Q$ values calculated from azimuthally averaged quantities,
\begin{equation}
Q=\frac{\langle \Omega \rangle \langle c_s \rangle}{\pi G \langle \Sigma \rangle},
\end{equation}
where $\langle \rangle$ means the azimuthal average. 
The panel shows $Q \sim 1 \propto r^{0}$ over almost all regions of the disk. 
This is a general feature of the self-gravitating disks without fragmentation and
the disk is in the quasi-steady state against gravitational instabilities. 

The bottom left panel shows the mid-plane gas temperature. The thick 
dashed line shows the theoretically estimated power law under the
assumption that local heating balances local 
cooling ($T \propto r^{-2.2}$; thick dashed line). 
This power law is calculated from (\ref{sec2_dynamics}), (\ref{sec2_qvalue}),
(\ref{sec2_cooling2}), and $\Omega \propto r^{-1.1}$. 
This power law does not agree with the disk structure because
the energy balance is not local, and the radial component of the 
radiative transfer heats the outer disk region.
We fitted the temperature to $T \propto r^{-1.1}$ (thick dashed-dotted line).
The power law $T \propto r^{-1.1}$ can be derived with the diffusion approximation 
(see \S \ref{section5_temp} for details).

The right top panel shows the profile of surface density $\Sigma$.
The thick solid lines show the power law that is predicted from (\ref{sec2_dynamics}), (\ref{sec2_qvalue}),
$\Omega \propto r^{-1.1}$, and $T \propto r^{-1.1}$. The thick dashed line is the power law  predicted from
equations (\ref{sec2_dynamics}), (\ref{sec2_qvalue}), $\Omega \propto r^{-1.1}$, and equation 
(\ref{sec2_cooling2}).
In the outer region with  $r>20$ AU, the surface 
density profile is consistent with the thick solid line. 
However, in the region of $r<20$ AU, the profile does not agree with the thick solid line 
because $Q$ is not constant (see left bottom panel),
and the assumption of $Q\propto r^0$ does not apply in this region. The thick-dashed line does
not agree with the profile over the entire region.

The right middle panel shows the profile for $\alpha$ parameter.
$\alpha$ parameter at each radius were calculated from
\begin{equation}
\langle \alpha \rangle_{t_0} \equiv \frac{1}{\Delta t_{\rm ave}}\int_{t_0}^{t_0+\Delta t_{\rm ave}} dt \langle T_{R \phi} \rangle  \left(\langle 
\frac{d\ln \Omega}{d \ln R} \rangle \langle \Sigma \rangle \langle c_s \rangle^2 \right)^{-1},
\end{equation}
where  $T_{R \phi}$ is the $R-\phi$ component of the viscous stress tensor.
The $\alpha$ parameter is averaged over $\Delta t_{\rm ave}$ where we took $\Delta t_{\rm ave}=2500$ years. 
The stress tensor associated with the gravitational field is given by
\begin{equation}
T_{R \phi}^{\rm grav}=\int dz \frac{g_R g_\phi}{4 \pi G},
\end{equation}
where $g_R$ and $g_\phi$ are the radial and azimuthal components of the gravitational force, respectively.
The stress tensor associated with  velocity field fluctuations or Reynolds stress is calculated by
\begin{equation}
T_{R \phi}^{\rm Reyn}=\Sigma \delta v_R \delta v_\phi.
\end{equation}
Here, we define a velocity fluctuation by $\delta \bmath{v}=\bmath{v}-\langle \bmath{v} \rangle$.
The total stress tensor is calculated from the sum of these tensors, 
$\langle T_{R \phi}\rangle=\langle T_{R \phi}^{\rm grav} \rangle+\langle T_{R \phi}^{\rm Reyn}\rangle$.

In the $\alpha$ profile, $t_0$ corresponds to
the epochs of Figs.~\ref{faceon_sigma}{\it f} (solid), {\it g} (dashed), and {\it h} (dotted).
The profile  can be described well by the theoretical estimate from
(\ref{sec2_dynamics}), (\ref{sec2_qvalue}), $T \propto r^{-1.1}$, and $\Omega \propto r^{-1.1}$
($\alpha \propto r^{1.65}$; thick solid line). As expected,  $\alpha$ was not constant.
This is a general feature of a non-isothermal self-gravitating disk.
Note that the model in which  local viscous heating balances local radiating cooling 
predicts a very steep radial profile  $\alpha \propto r^{3.3}$ (thick dashed line) and fails to explain the $\alpha$ 
profile obtained from the numerical simulation.




The bottom right panel shows the vertical optical depth, $\tau_z$.
The optical depth was calculated 
from $\tau_z=\int^{\infty}_{\infty} \kappa \rho dz$. In almost 
all regions of the disk, the vertical optical depth is greater than unity, $\tau_z>1$.
Again, the thick solid line describes the radial profile very well.

\begin{figure}
\includegraphics[width=90mm]{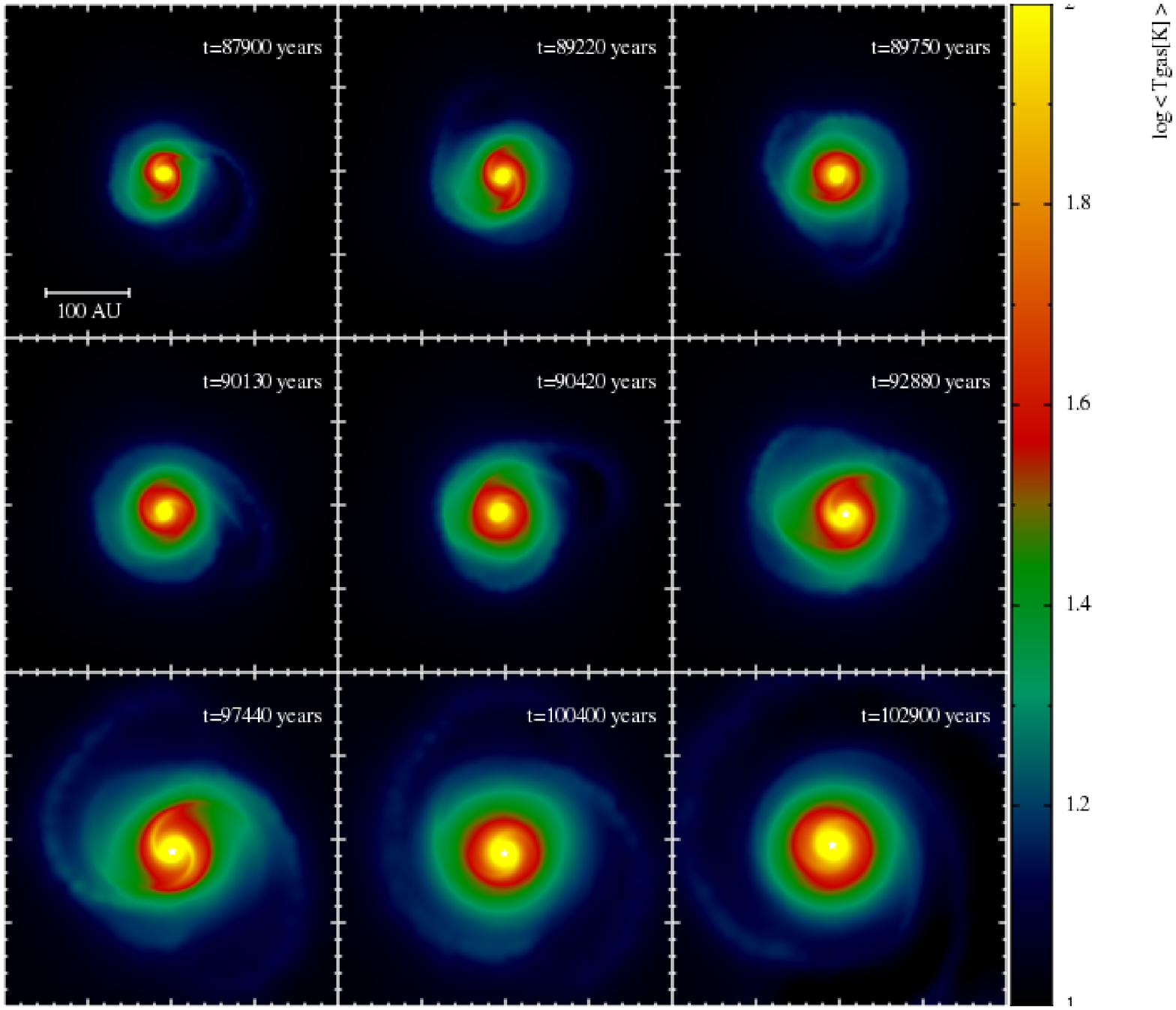}
\caption{
Time sequence for the density-weighted gas temperature at the cloud center, viewed face-on for model 1.
Elapsed time after the cloud core begins to collapse is shown in each panel. 
}
\label{faceon_Tgas}
\end{figure}

\begin{figure}
\includegraphics[width=70mm,angle=-90]{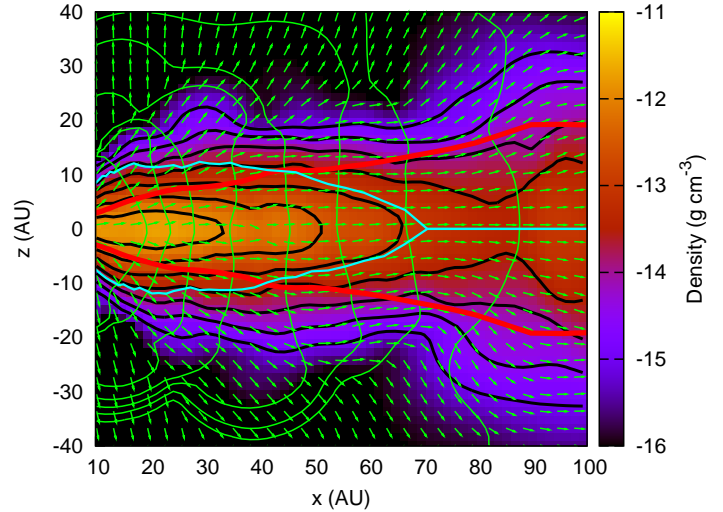}
\caption{
Density contour on $y=0$ plane at the same epoch as in Fig.~\ref{faceon_sigma}{\it i} of Model 1. 
Red lines show scale height of disk, $H(r)$;
cyan lines show height of vertical photosphere;
green lines and arrows show temperature contours and direction of radiation flux, respectively.
}
\label{rz_map}
\end{figure}

\subsubsection{Cooling time of disk}
In Fig. \ref{tcool_profile}, we show the azimuthally averaged cooling time of the disk normalized by 
orbital frequency, $\Omega t_{\rm cool}$. Here, cooling time is calculated as
\begin{eqnarray}
\label{def_mass}
t_{\rm cool}=\frac{\Sigma c_s^2}{\gamma-1}\frac{1}{2 \sigma T_{\rm mid}^4}(\tau+\frac{1}{\tau}),
\end{eqnarray}
where $T_{\rm mid}$ is mid-plane temperature and $\tau=\int^{\infty}_{0} \kappa \rho dz$. 
We average above $t_{\rm cool}$ in the azimuthal direction.
In outer region of the disk, the cooling time satisfies $\Omega t_{\rm cool} \lesssim 1$ and 
it is expected that the gas behaves almost isothermally during GI growing.
Note that, for our disk,  this cooling time is not the timescale on 
which disk temperature decreases because the temperature of
our disk is largely affected by radial radiative transfer. 
However, we use this cooling time as
an indicator how the disk gas behaves against the dynamical compression by GI. When the cooling time is
sufficiently small (large), the gas evolves isothermally (adiabatically) during GI growing.
As we have seen in Fig. \ref{radial_profile}, $Q$ value of the disk is close to 1.
Thus, our disk apparently  satisfies the fragmentation criterion 
suggested by \citet{2001ApJ...553..174G} and \citet{2003MNRAS.339.1025R}.
However, the disk does not fragment.
This shows that the fragmentation criterion based on disk 
cooling time is not sufficient condition for fragmentation.

\subsection{Discussion}
\label{self_grav_disk_discussion}

\subsubsection{Temperature profile determined by radial radiative transfer in an optically thick disk}
\label{section5_temp}
\begin{figure*}
\includegraphics[width=50mm,angle=-90]{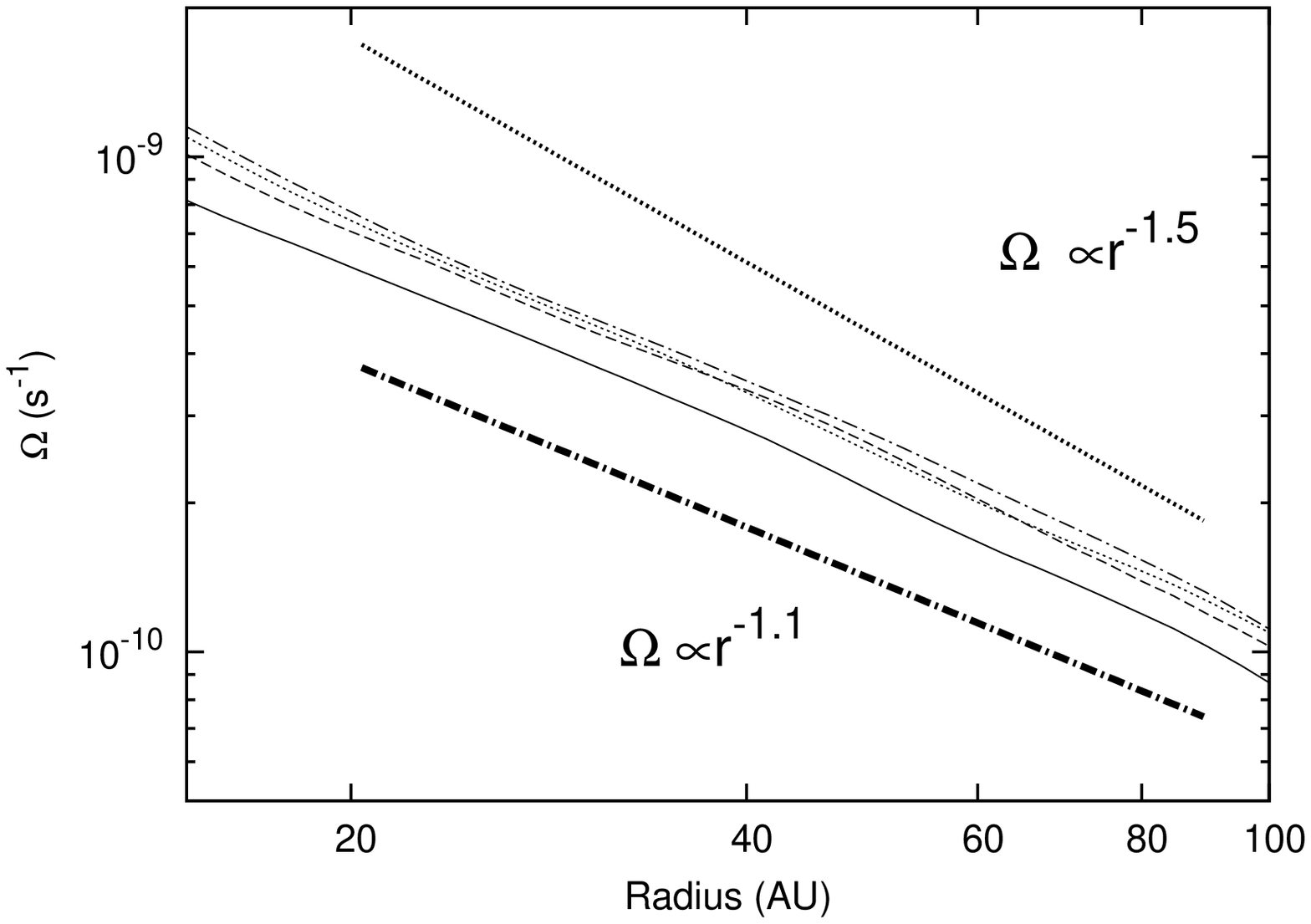} 
\includegraphics[width=50mm,angle=-90]{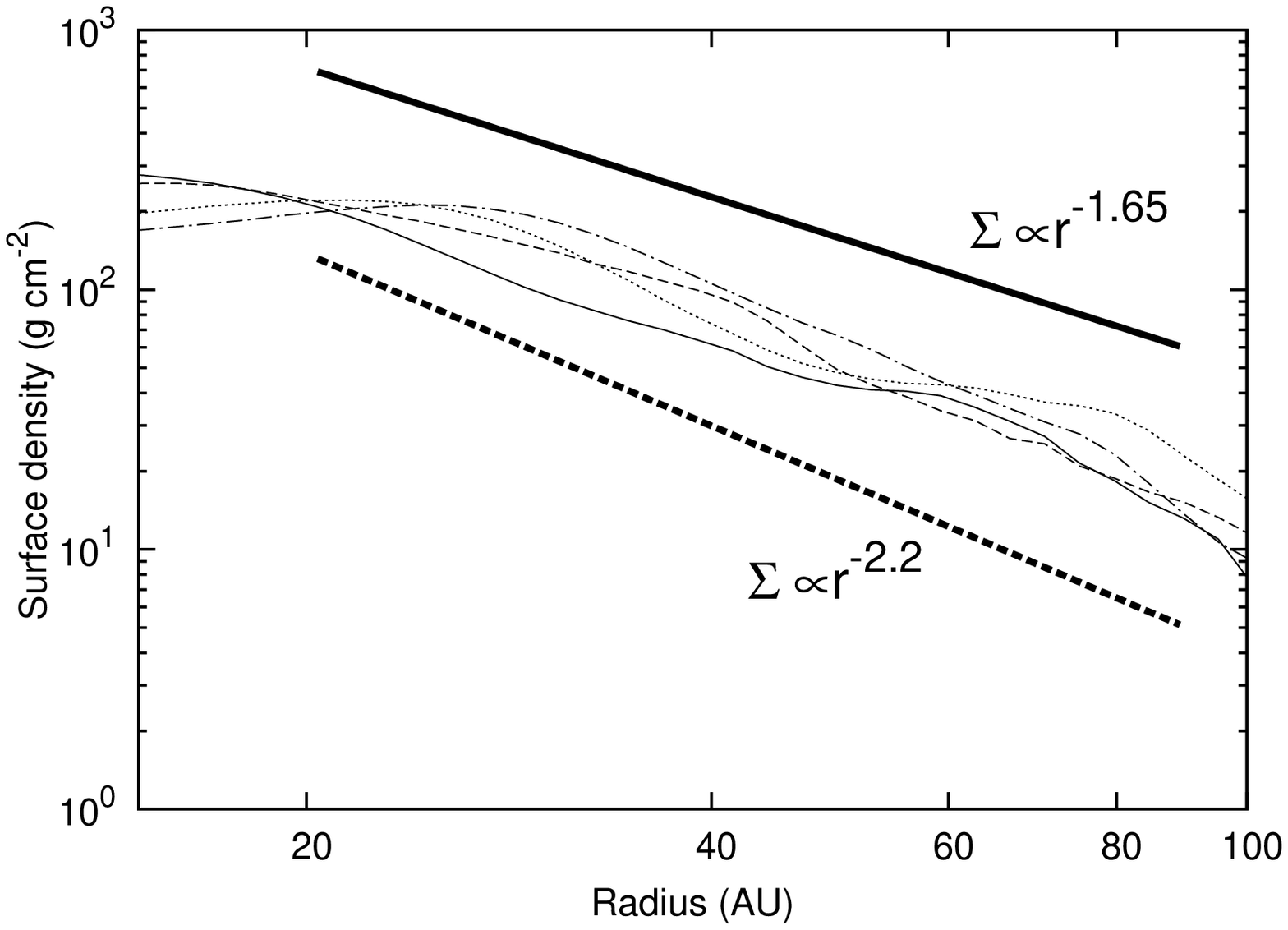}
\includegraphics[width=50mm,angle=-90]{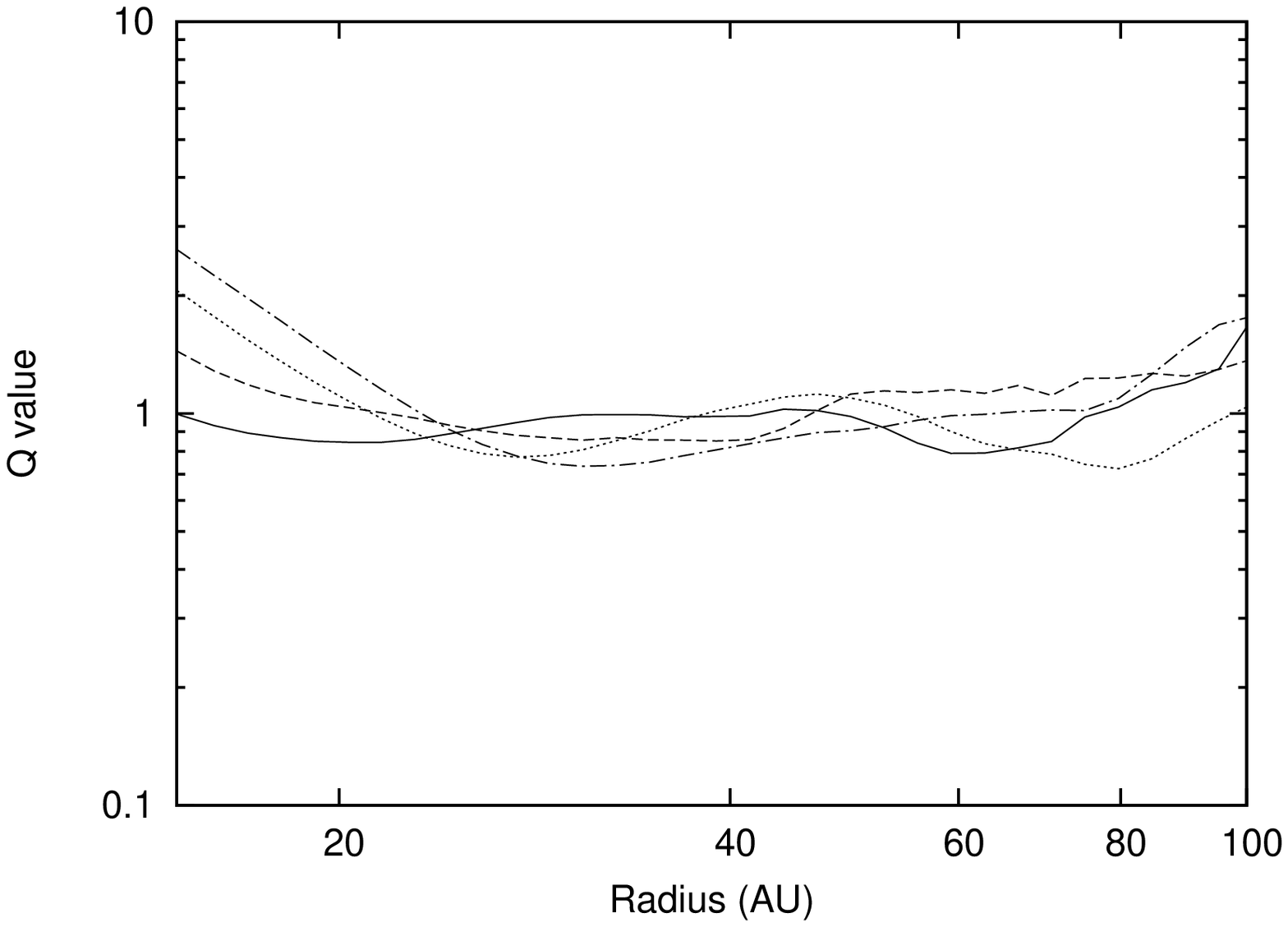}
\includegraphics[width=50mm,angle=-90]{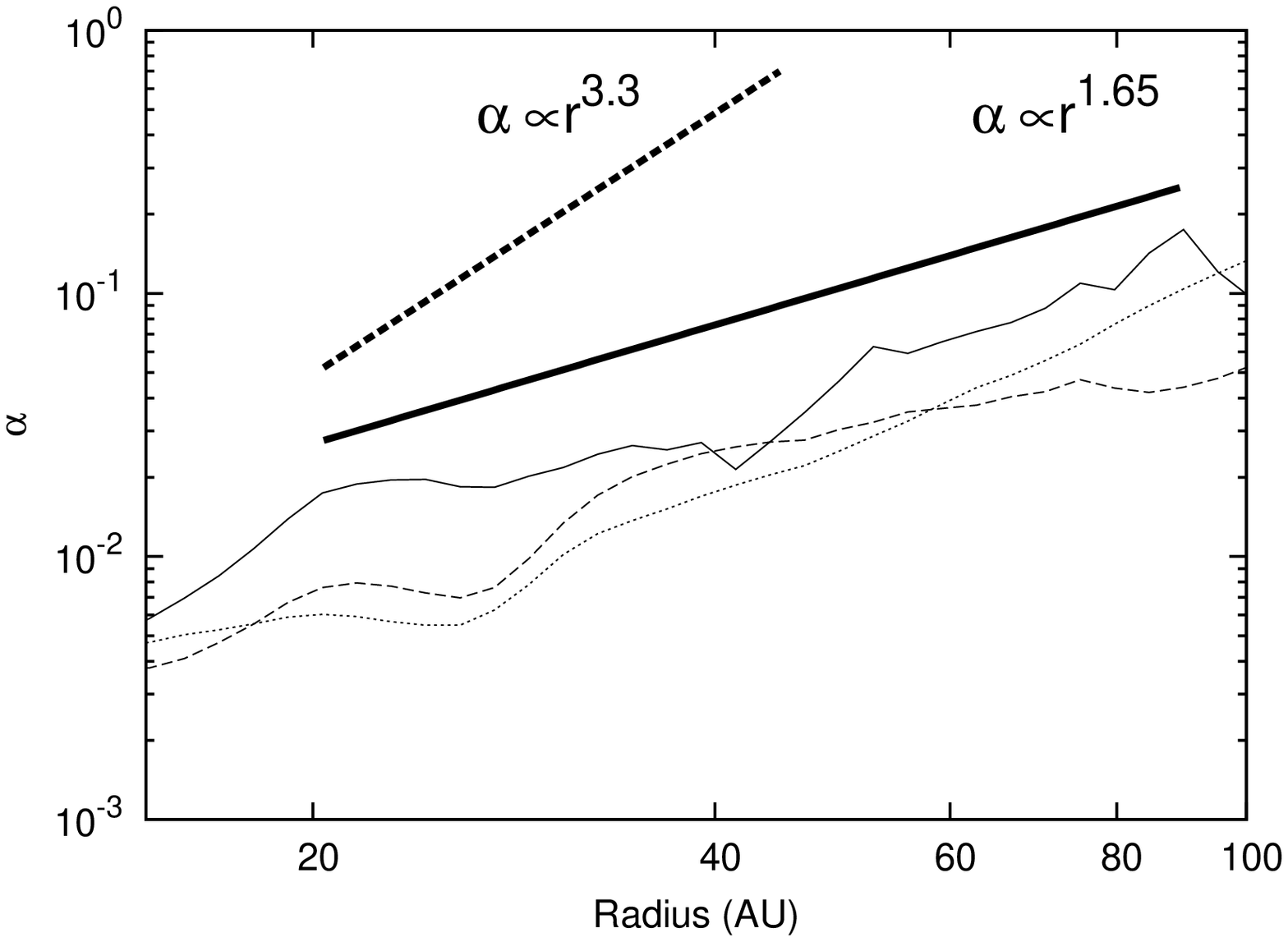}
\includegraphics[width=50mm,angle=-90]{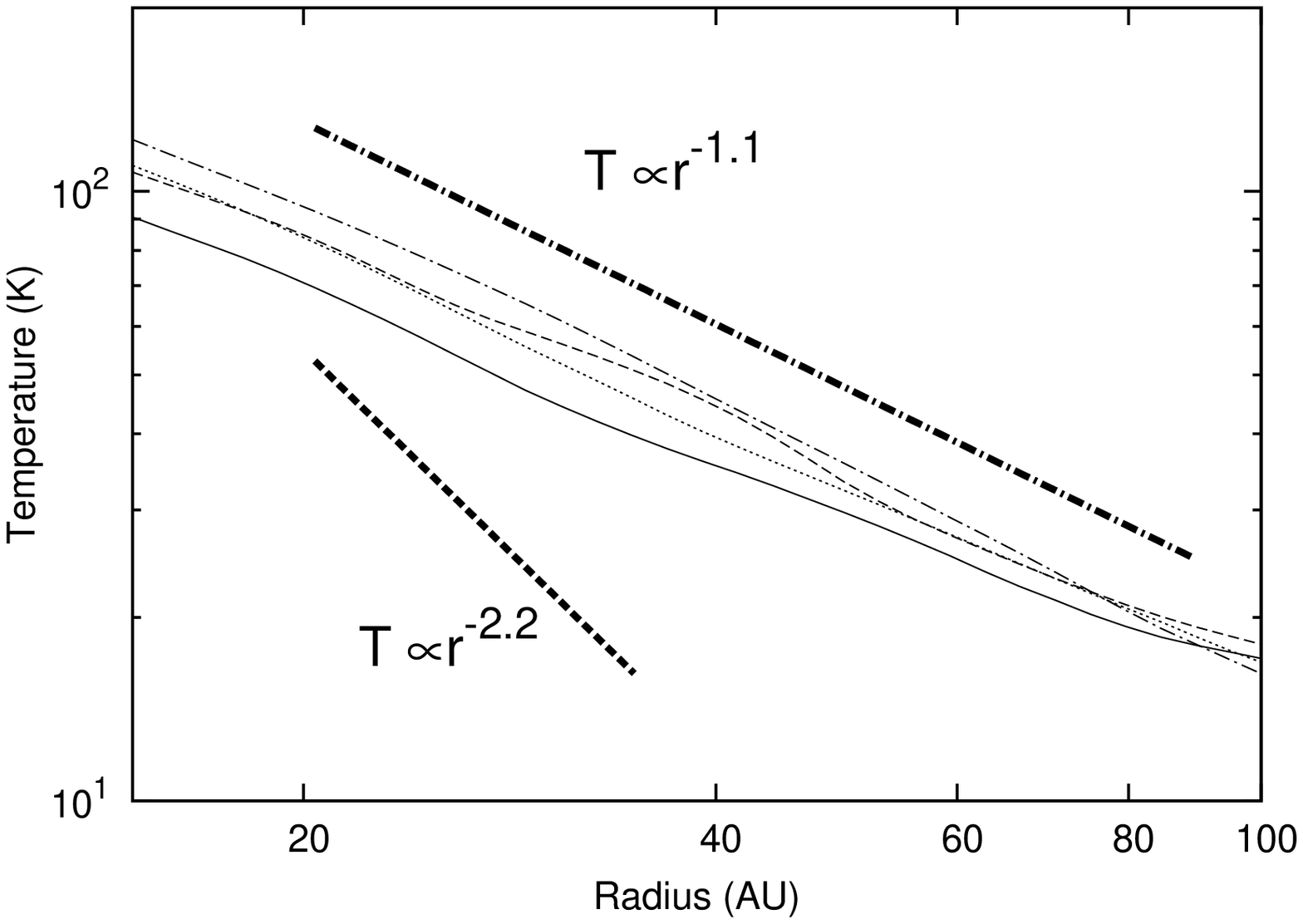}
\includegraphics[width=50mm,angle=-90]{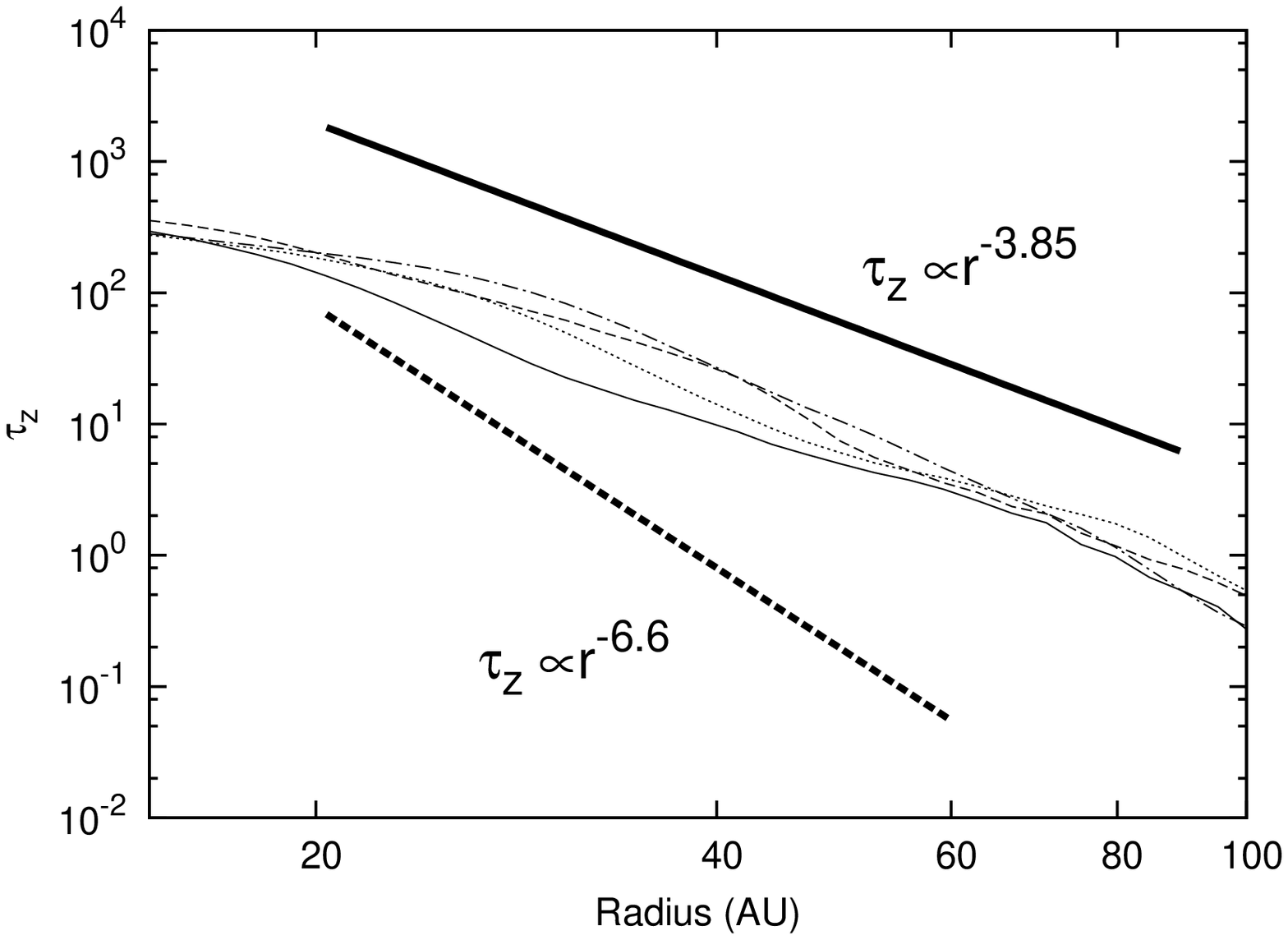}

\caption{
Azimuthally averaged radial profiles of the disk of Model 1.
Thin solid, dashed, dotted, and dashed-dotted lines correspond to profiles at the same epoch as in 
Fig.~\ref{faceon_sigma}{\it f}, {\it g}, {\it h}, and {\it i}, respectively.
Top, middle, and bottom left panels show the angular velocity, Toomre's $Q$ value, and gas temperature, respectively.
Top, middle, and bottom right panels show the surface density, $\alpha$ parameter, and vertical optical depth, respectively.
Thick solid lines show power law profiles estimated 
from (\ref{sec2_dynamics}), (\ref{sec2_qvalue}), $T \propto r^{-1.1}$, and $\Omega \propto r^{-1.1}$. 
Thick dashed lines show power law profiles estimated 
from  (\ref{sec2_dynamics}), (\ref{sec2_qvalue}), (\ref{sec2_cooling2}), and $\Omega \propto r^{-1.1}$. 
Thick dotted lines shows power law of Kepler rotation $\Omega\propto r^{-3/2}$.
Thick dashed-dotted lines are fittings used for theoretical estimates.
}
\label{radial_profile}
\end{figure*}

The results in \S \ref{results_disk} shows that the temperature profile of the disk formed in our simulation is
$T \propto r^{-1.1}$. This is significantly shallower than the disk in which local
heating balances local cooling $T\propto r^{-2.2}$ (see, Fig. \ref{radial_profile}) 
and steeper than the passively irradiated disk $T\propto r^{-\frac{3}{7}}$. 
In this subsection, we analytically derive the profile $T\propto r^{-1.1}$.

The right bottom panel in  Fig. 
\ref{radial_profile} shows that the vertical optical depth is greater than unity over almost the entire region.
Thus, we can use the diffusion approximation to derive the disk temperature profile.
Under the assumptions that disk is steady, viscous heating is negligible, and that the 
radiation temperature is equal to the gas temperature, the energy equation can be expressed as
\begin{equation}
\label{flux}
\nabla \cdot \mathbf{F}_{\bm r}=0.
\end{equation}
Figure \ref{faceon_Tgas} and \ref{rz_map} shows that the temperature distribution can be regarded as 
axisymmetric and vertically isothermal.
Thus, we can assume that $F_r \gg F_z,~F_\phi$.
Under the assumption that $F_r \gg F_z,~F_\phi$, (\ref{flux}) becomes
\begin{equation}
\label{flux_fr}
\frac{1}{r}\frac{ \partial (r F_r)}{\partial r} =0.
\end{equation}
Around the mid-plane, the diffusion approximation is justified, and (\ref{flux_fr}) becomes
\begin{equation}
r \frac{\sigma T^3}{\kappa \rho}\frac{\partial T}{\partial r}={\rm const.}.
\end{equation}
Using the relation $ \Sigma=\rho c_s/\Omega $, we have
\begin{equation}
\label{flux_n}
2.5~ n_T - n_\Sigma -n_\Omega =0.
\end{equation}
Using (\ref{sec2_dynamics}), (\ref{sec2_qvalue}), $n_\Omega=-1.1$ and (\ref{flux_n}), we obtain
\begin{equation}
\begin{split}
\alpha\propto r^{1.65},~\Sigma\propto r^{-1.65},~T\propto r^{-1.1}.
\end{split}
\end{equation}
This agrees very well with the simulation results (see, the thin lines (simulation results)
and  thick solid lines (theoretical estimate) in Fig. \ref{radial_profile}).

\subsubsection{Importance of non-local radiative transfer}
As pointed out in \S \ref{intro}, the locality of radiative cooling has been
assumed in many previous simulations 
\citep[e.g.,][]{2003MNRAS.339.1025R,2005MNRAS.358.1489L,2005MNRAS.364L..56R,2012MNRAS.420.1640R,
2011MNRAS.416L..65P}
and in many analytical studies \citep{2005ApJ...621L..69R,2009MNRAS.396.1066C,2011ApJ...740....1K} 
to investigate the nature of GI and the conditions of disk fragmentation with radiative cooling.
In contrast, our simulations show that radiation can
transfer significant energy in the radial direction even in the absence of the stellar irradiation. 

Consequently, the temperature profile becomes $T \propto r^{-1.1}$ in our simulation
and is significantly shallower than $T \propto r^{-2.2}$, which is expected under the assumption that local viscous
heating balances local radiative cooling.
Thus, the local energy balance between radiation and viscous heating is not satisfied in the disk formed
in our simulations.
This indicates that the assumption of local radiative cooling is not necessarily valid in a
massive disk around a low mass star. 
The radial profiles of the surface density and the viscous $\alpha$ parameter are
consistent with a self-gravitating quasi-steady disk model with a given rotation profile
and $T \propto r^{-1.1}$. Thus, the disk is in the quasi-steady state with an energy balance that is not local.

Therefore, we conclude that the local treatment of radiative cooling is not suitable approximation to investigate massive disk
around low mass star.

We only calculated disk evolution just about $ 10^4$ years after protostar formation and the disk and the protostar
were still embedded in a massive envelope (Class 0 phase).
One might think that non-local radiative transfer does not play the role in later evolutionary phase. 
However, with two-dimensional radiation hydrodynamics simulation, \citet{1993ApJ...411..274Y} showed 
that an almost isothermal temperature distribution in vertical direction also realize in the late evolutionary phase of disk
at which 95 \% of initial cloud core has already accreted onto the protostar or disk.
Thus, we expect that non-local radiative transfer also plays an important role for the temperature structure of disks
in the later evolutionary phase.
Note that a large simulation box in which the boundary is far from disk photosphere
is crucial to investigate the effects of non-local radiative transfer, otherwise
radiation flux is easily affected by the boundary conditions.

We could not find evidence that the non-local energy transport of GI suggested by 
\citet{1999ApJ...521..650B} plays an important role because GI heating is itself 
small compared to radiation energy transfer in the outer region of the disk.

\subsubsection{Applicability of fragmentation criterion based on disk cooling time}
\label{application_fragmentation_criterion}
\begin{figure}
\includegraphics[width=50mm,angle=-90]{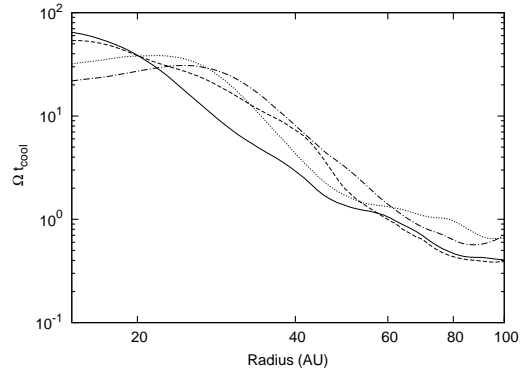} 

\caption{
Azimuthally averaged $\Omega t_{\rm cool}$, where $t_{\rm cool}$ is the cooling time of the disk.
Thin solid, dashed, dotted, and dashed-dotted lines correspond to profiles at the same epoch as in 
Fig.~\ref{faceon_sigma}{\it f}, {\it g}, {\it h}, and {\it i}, respectively.
}
\label{tcool_profile}
\end{figure}

In Fig. \ref{tcool_profile}, we show that the cooling time of disk corresponds to $\Omega t_{\rm cool}\lesssim 1$ in outer region
and our disk apparently satisfies the fragmentation criterion based on disk cooling time ($Q \sim 1$ and $\Omega t_{\rm cool}\sim 1$).
Because the cooling time is very short compared to the orbital period, the gas 
evolves almost isothermally during GI growing in outer region. 
However, the disk does not fragment. 

In our disk, the most unstable wave-length of GI is relatively large and the global spiral arms develop.
The spiral arms readjust the surface density and the disk is stabilized. 
Such a readjustment is also observed in previous works \citep{1994ApJ...436..335L,2010ApJ...708.1585K}. Especially,
\citet{2010ApJ...708.1585K} shows that the disk fragmentation is suppressed by the readjustment even with isothermal 
equation of state. 
Because such a stabilizing process also play the role in the realistic disk, 
we conclude that  at least the fragmentation criterion is not sufficient condition for 
fragmentation in realistic disk around low mass star.

A significant difference between our disk and disks used in the previous works with local cooling law
is the most unstable wave-length of GI. 
The disks used in previous studies using local cooling law
\citep[e.g.,][]{2003MNRAS.339.1025R,2012MNRAS.427.2022M} have 
very small value of the most unstable wave-length 
and the spiral arms formed in these simulations
are not geometrically thick and have large azimuthal mode numbers $m$. It is expected that the efficiency of the readjustment
promoted by such spiral arms is small compared to that promoted by global spiral arms which may form in realistic disk.

Is the fragmentation criterion based on local cooling  a necessary 
condition for disk fragmentation ?
We think this is not true, either.
Although the efficiency of the radiative cooling may affect the evolution of condensations formed by GI,
cooling criterion is not necessary for the fragmentation of $Q \sim1$ disk.
This can be understood from the previous works that employed the simplified equation of state.
For example, \citet{2006ApJ...650..956V}, \citet{2010ApJ...718L..58I}, \citet{2011MNRAS.416..591T}, and 
\citet{2011MNRAS.413.2767M}  employed barotropic 
equation of state and the gas evolves adiabatically 
when $\rho \gtrsim 10^{-13} \cm$ (the condensation exceeds this density at
very early phase of its evolution). Even with such a stiff (or adiabatic) equation of state,
fragmentation is observed. 

Therefore, we conclude that, in general,  the fragmentation 
criterion based on disk cooling time 
($Q \sim 1$ and $\Omega t_{\rm cool}\sim 1$) 
is  not necessary nor sufficient conditions for fragmentation.

One may think that our results appear very different from 
the previous results which
show the disk fragmentation occurs 
when it satisfies the fragmentation criterion 
and does not occur when it does not satisfy the 
criterion \citep[e.g.,][]{2005MNRAS.364L..56R,2008A&A...480..879S,2009MNRAS.392..413S}.
However, we emphasize that there is no contradiction.
Logically speaking, there are four possible cases,
\begin{enumerate}
\item The disk satisfies ($Q \sim 1$ and $\Omega t_{\rm cool}\sim 1$) and the disk fragments,
\item The disk does not satisfies ($Q \sim 1$ and $\Omega t_{\rm cool}\sim 1$) and the disk does not fragment,
\item The disk satisfies ($Q \sim 1$ and $\Omega t_{\rm cool}\sim 1$) but the disk does not fragment,
\item The disk does not satisfies ($Q \sim 1$ and $\Omega t_{\rm cool}\sim 1$) but the disk fragments.
\end{enumerate}
The previous works mentioned above show the cases of (i) and (ii). 
On the other hand, we point out there exist the cases of (iii) and (iv).
Therefore, there is no contradiction.

Note, however, that
the finite number of examples of (i) and (ii) are 
not enough to prove the statement that
``ALL the disk fragments when it satisfies 
$Q \sim 1$ and $\Omega t_{\rm cool}\sim 1$"
or ``ALL the fragmenting disk must satisfy
$Q \sim 1$ and $\Omega t_{\rm cool}\sim 1$"
because we cannot prove non-existence of the exceptions with finite
number of the simulations.
On the other hand, 
just one counter-example is enough to reject these statements.
In this paper, we show there exists (iii) case 
and this is a counter-example for the statement,
``ALL the disk fragments when it satisfies 
$Q \sim 1$ and $\Omega t_{\rm cool}\sim 1$".
Therefore, the fragmentation criterion is not sufficient condition.
On the other hand, the previous works we mentioned above
show that there exists (iv) case. This is a counter-example for the statement
``ALL the fragmenting disk must satisfy 
$Q \sim 1$ and $\Omega t_{\rm cool}\sim 1$".
Therefore, the fragmentation criterion is not necessary condition.
Have we emphasize that the discussions above is based on the azimuthally 
averaged quantities (see \S \ref{diff_frag}).

\subsubsection{Resolution consideration for disk fragmentation simulations}
\label{sec5_res}

\subsubsection*{Resolution consideration of the simulations performed in this paper}

\begin{figure*}
\includegraphics[width=50mm,angle=-90]{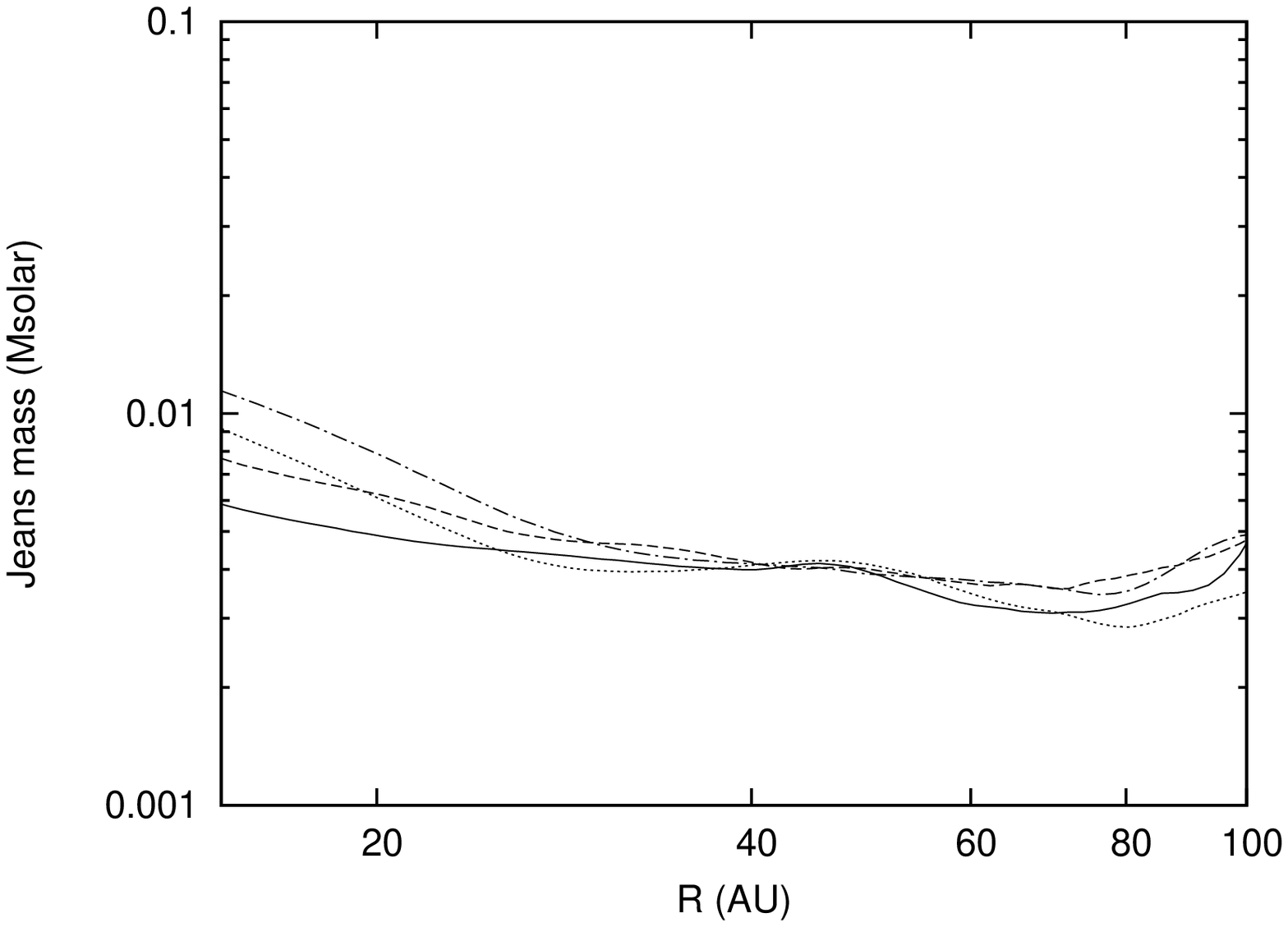}
\includegraphics[width=50mm,angle=-90]{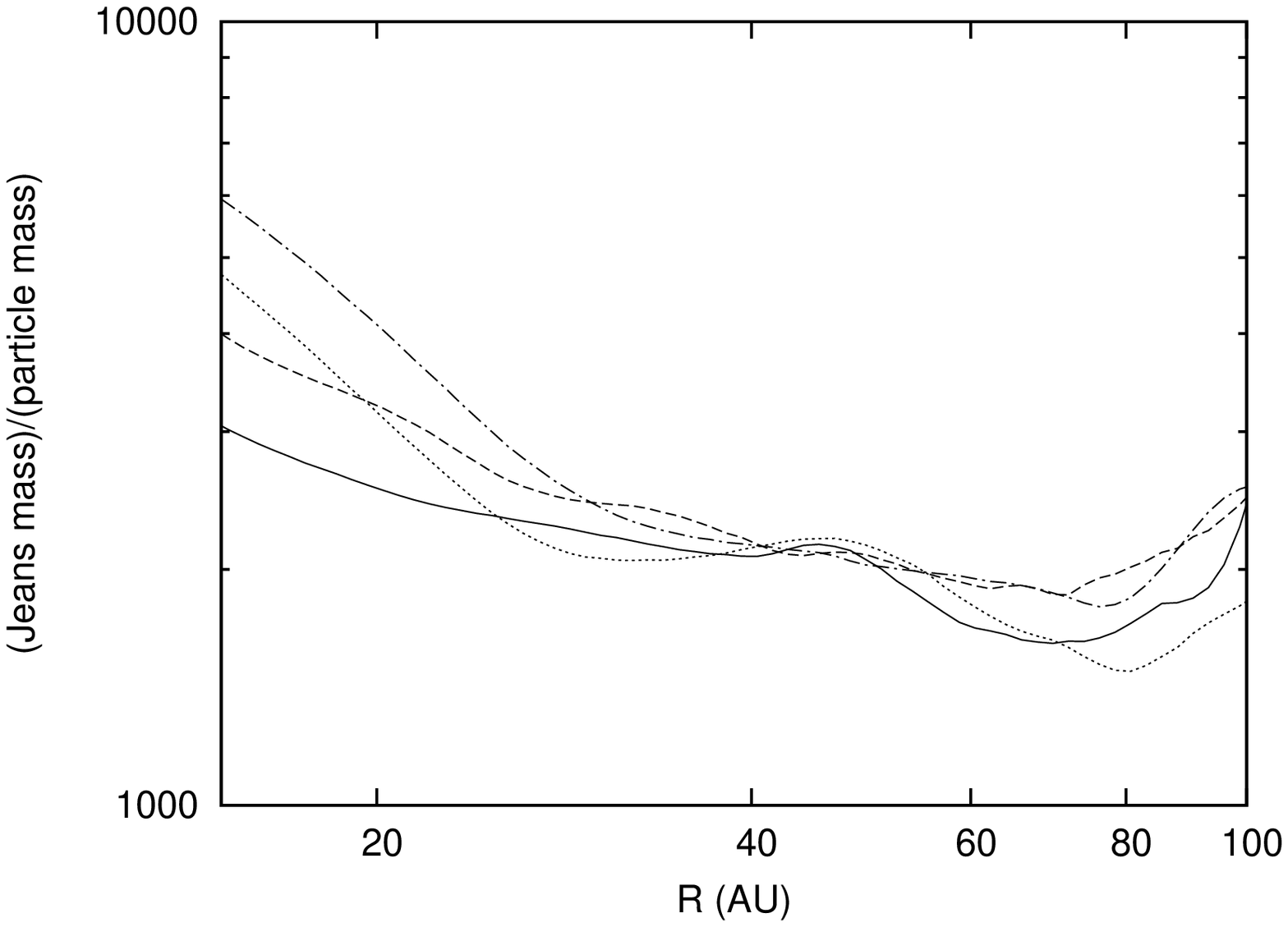}
\includegraphics[width=50mm,angle=-90]{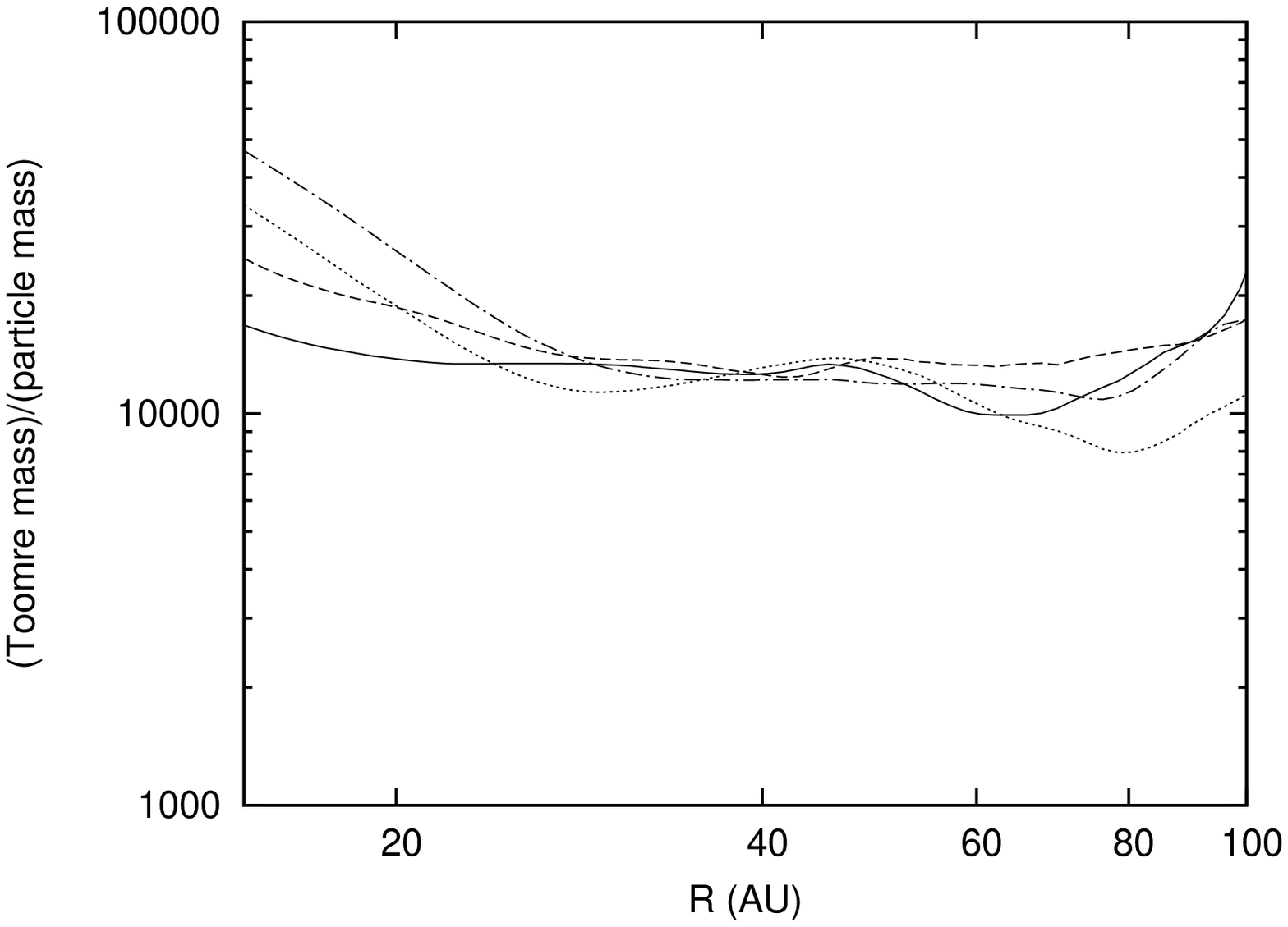}
\includegraphics[width=50mm,angle=-90]{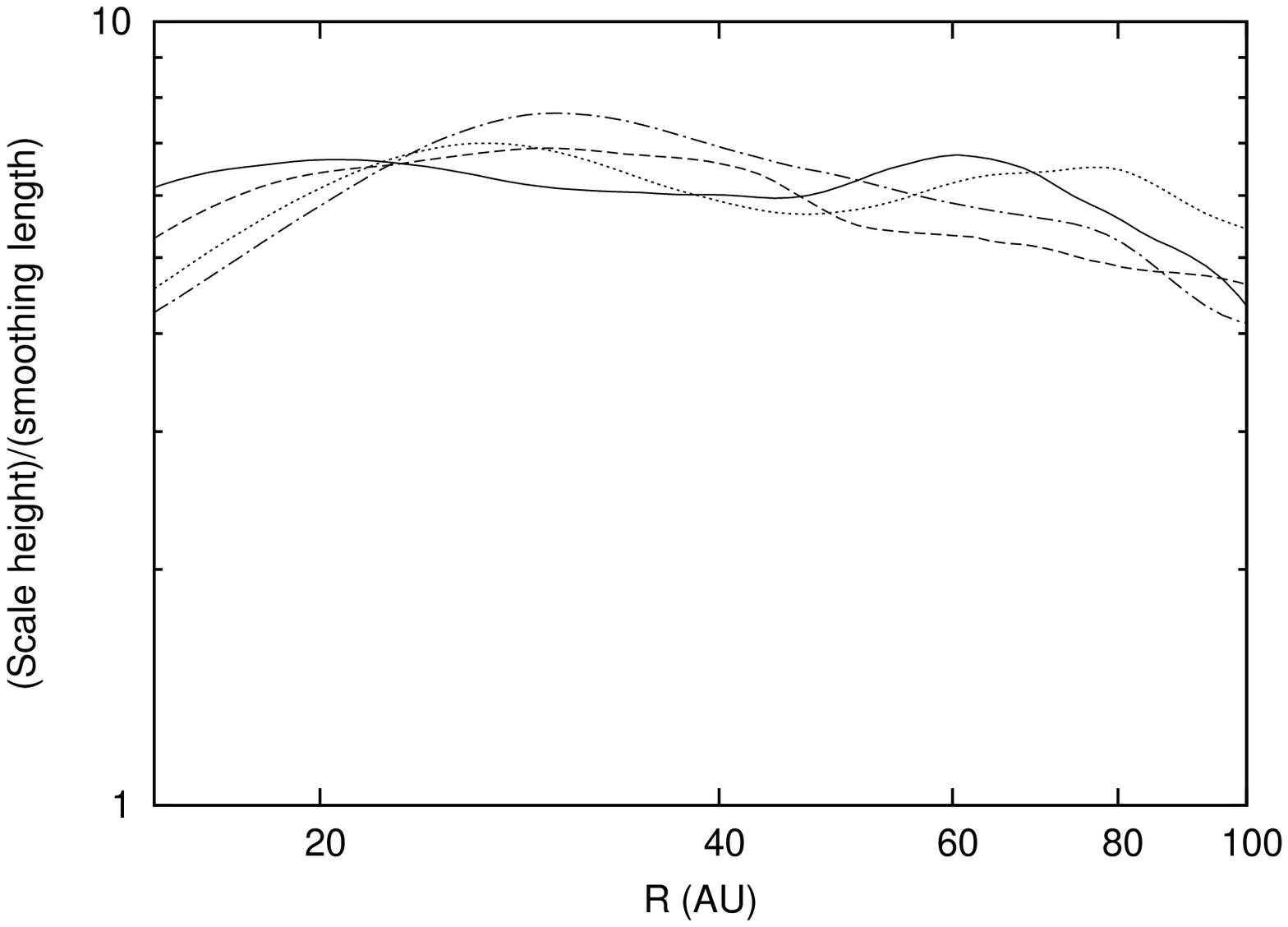}

\caption{
Azimuthally averaged Jeans mass (top left), ratio of Jeans mass to SPH particle mass (top right),
ratio of Toomre mass to SPH particle mass (bottom left) and ratio of scale height to smoothing length (bottom right).
Each line corresponds to the same epoch as in Fig.~\ref{faceon_sigma}{\it f} 
(solid), {\it g} (dashed), {\it h} (dotted), and {\it i} (dashed-dotted).
\label{resolution1}
}
\end{figure*}

Numerical resolution is considered to be an important factor in simulations of disk fragmentation.
Recently, \citet{2012MNRAS.427.2022M} argued that the condition for disk fragmentation 
using local cooling law strongly depends on numerical resolution. 
Even though our simulation is based on a more realistic radiative transfer method and the disk structure is
completely different from theirs,
it is important to check whether the numerical resolution in our simulation was sufficient.
As \citet{1997MNRAS.288.1060B} discussed, the SPH method cannot correctly simulate the fragmentation phenomenon
if the minimum resolvable mass $M_{\rm min}=2 N_{n} m_{p}$ is greater
than the Jeans mass $M_{\rm Jeans}$, where $N_{n}$ and $m_{p}$ are the number of neighbors
and  mass of the SPH particles, respectively.
Here, according to \citet{1997MNRAS.288.1060B}, the Jeans mass is given by
\begin{equation}
\label{eq_Jeans_mass}
M_{\rm Jeans} = \left( \frac{5 R_g T}{2 G \mu} \right)^{3/2} \left(\frac{4 \pi}{3} \rho \right)^{-1/2},
\end{equation}
where $R_g$ is the gas constant.
 On the other hand, \citet{2006MNRAS.373.1039N} suggested that Toomre mass,
\begin{equation}
\label{eq_Jeans_mass}
M_{\rm Toomre} = \frac{\pi c_s^4}{G^2 \Sigma},
\end{equation}
is more appropriate mass scale for fragmentation in disk 
and it should be resolved with $6 N_n$.
Furthermore he also suggested that the disk vertical scale height should be resolved at least
$4 h_{\rm sml}$, where $h_{\rm sml}$ is the smoothing length. 
In Fig. \ref{resolution1}, we plotted azimuthally averaged Jeans mass (top left), $M_{\rm Jeans}/m_{p}$ (top right),
$M_{\rm Toomre}/m_{p}$ (bottom left) and $ H/h_{\rm sml} $ (bottom right)
for the same epochs as in Fig.~\ref{faceon_sigma}{\it f} (solid), {\it g} (dashed), {\it h} 
(dotted), and {\it i} (dashed-dotted).

The Jeans mass of our disk is about $4 \times 10^{-3}~M_\odot$ over almost the entire region of the disk;
this value is consistent with the initial clump mass that formed 
in model 2. As shown in top right panel, the Jeans mass is resolved 
by larger than  $\sim 2000$ particles. In our simulations, we adopted the number of neighbors to be
$N_{n}\sim 50$; hence, the mass resolution
is about 20 times higher than that required by \citet{1997MNRAS.288.1060B}.

The Toomre mass is resolved by larger than $\sim 10000$ particles and this
is about 30 times higher than the resolution criterion suggested by \citet{2006MNRAS.373.1039N}.
The bottom right panel shows that the vertical scale height $H$ is resolved by larger than $4 h_{\rm sml}$ and
our simulation also satisfies the resolution requirement for vertical scale height.
Therefore, we conclude that the numerical resolution in our simulation 
was sufficient to resolve fragmentation in the disk.

\subsubsection*{Resolution consideration of the simulations with local cooling law}
Why does the disk fragmentation criterion with local cooling law strongly depend on 
numerical resolution ? To answer this question, we investigate the resolution requirement of
the disk used in \citet{2005MNRAS.364L..56R} and \citet{2012MNRAS.427.2022M} with the quasi-steady state structure.
In \S 2, we investigated the steady state of the self-gravitating disk 
with local cooling law. As we pointed out,
the temperatures of the disks used in \citet{2005MNRAS.364L..56R,2012MNRAS.427.2022M} 
are very small in the quasi-steady state. Therefore, we can  expect that
the requirement on numerical resolution for their disks which settled into quasi-steady states is very severe.
In this subsection, we only consider the resolution requirement suggested by \citet{1997MNRAS.288.1060B}.

\begin{figure}

\includegraphics[width=50mm,angle=-90]{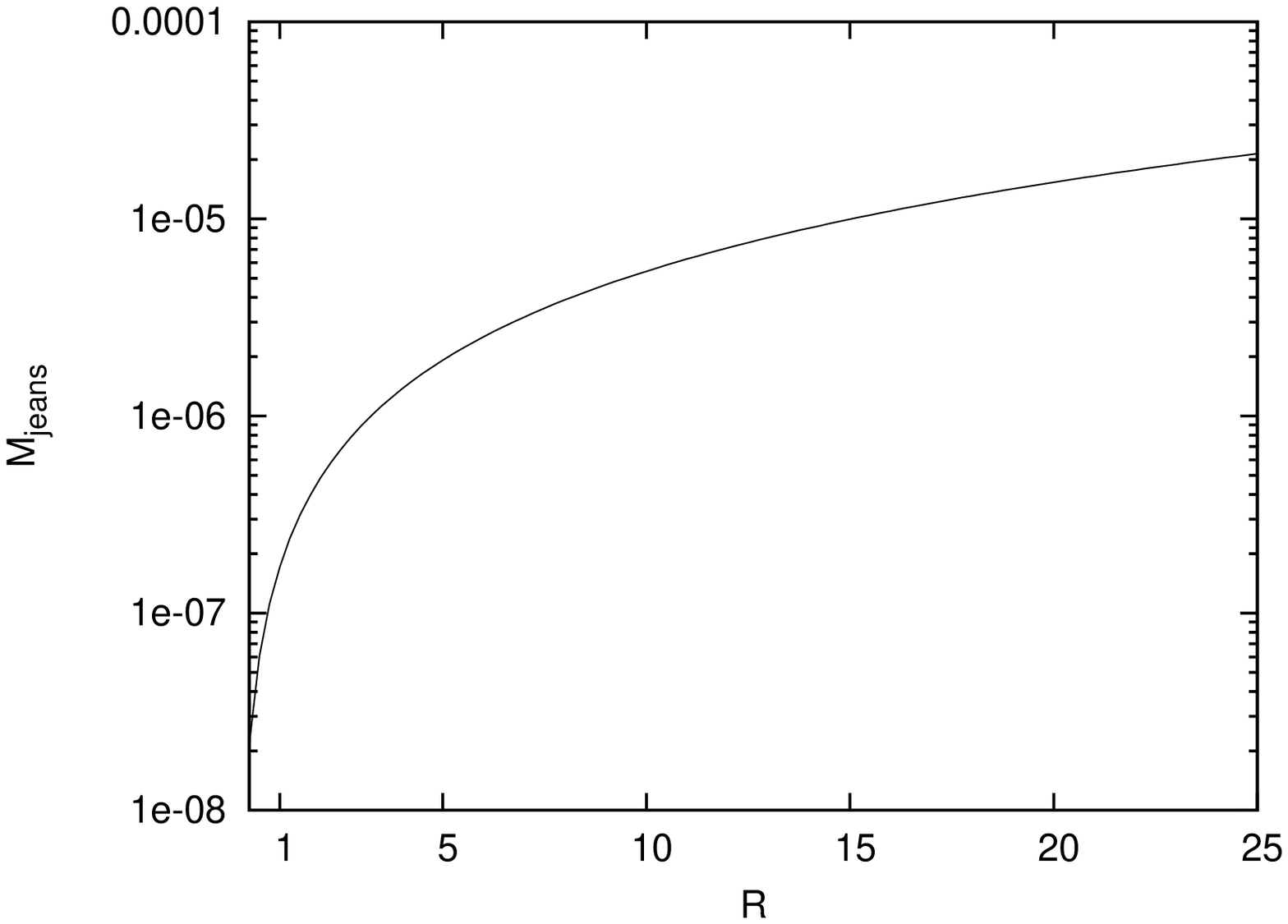}
\includegraphics[width=50mm,angle=-90]{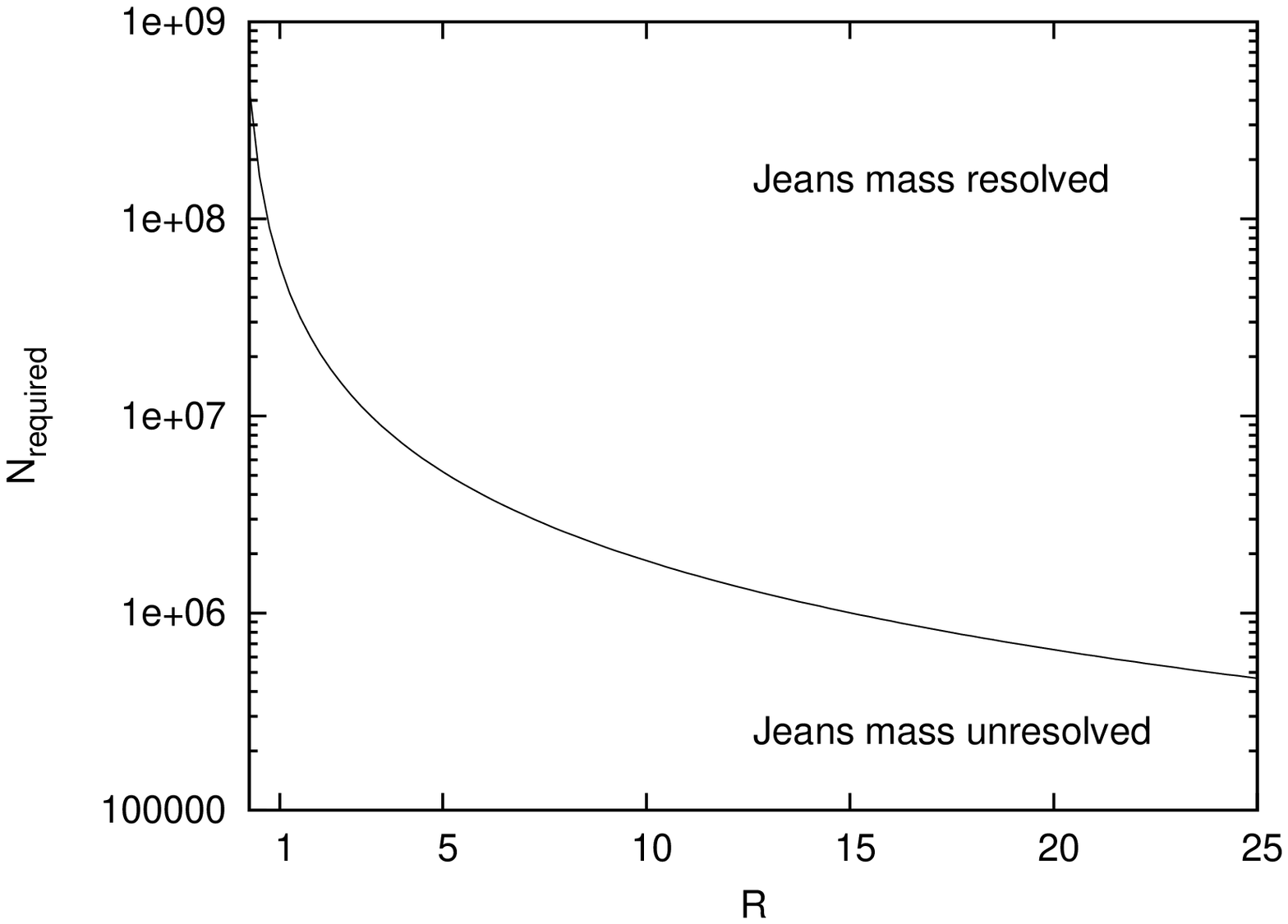}

\caption{
Jeans mass (top) and required number of particles to resolve the Jeans mass,
$N_{\rm req}=2 N_{n} M_{\rm disk}/M_{\rm Jeans}$ (bottom) for a disk in which 
$M_{\rm star}=1,~M_{\rm disk}=0.1,~r_{\rm in}=0.25,~r_{\rm out}=25$ are adopted with the disk 
in the quasi-steady state with local cooling law (\ref{sec2_gammie}).
\label{resolution2}
}
\end{figure}

In Fig. \ref{resolution2}, we show the Jeans mass (top) and the required 
particle number to resolve the Jeans mass, 
$N_{\rm req}=2 N_{n} M_{\rm disk}/M_{\rm Jeans}$ (bottom) as functions of the radius
for a disk that has dimensionless parameters of
$M_{\rm star}=1,~M_{\rm disk}=0.1,~r_{\rm in}=0.25,~r_{\rm out}=25$ and is 
in the quasi-steady state $\Sigma \propto r^{-\frac{3}{2}},~T \propto r^{0}, \alpha\propto r^{0}$.
Here according to \citet{2012MNRAS.420.1640R}, we approximated 
$M_{\rm Jeans}=\pi \Sigma H^2$ for comparison and we assumed $N_{n}=50$.
The Jeans mass is $M_{\rm Jeans} \lesssim 10^{-5} $. This corresponds to
about 0.01 $M_{\rm Jupiter}$ if we regard the 
central star mass as $1 ~M_{\odot}$. This value of the Jeans mass is very small compared to that in our simulation and 
to the mass of wide orbit planets ($\sim 10~M_{\rm Jupiter}$). 

The bottom panel in Fig. \ref{resolution2} shows that a particle number of $N_{p} \gtrsim 10^7$ 
is required to resolve the Jeans mass.
Note that, with $N_{p}<500,000$, the Jeans mass is not resolved over the entire disk region.
\citet{2012MNRAS.427.2022M} showed that the convergence of the calculation around 
$N_{p} \gtrsim10^7$, and their results are consistent with our estimate. 
Our estimate is more severe than that of \citet{2012MNRAS.420.1640R} in the outer disk region.
The difference is because \citet{2012MNRAS.420.1640R} assumed $\Sigma \propto r^{-1}$
instead of assuming the steady state as in (\ref{sec2_dynamics}).
Therefore, they derived $T\propto r^{1/2}$ and found a larger $M_{\rm Jeans}$ in the outer disk region. Note, however, that
such a disk is not realized with local cooling law because the surface density profile also changes
because of the angular momentum transfer to realize the quasi-steady 
state structure \citep[see Fig. 1 of][]{2011MNRAS.416.1971B}.

Note also that the resolution requirement is estimated using the initial total disk mass and initial
cutoff radii, $r_{\rm in}$ and $r_{\rm out}$.
As the disk evolves, gas accretes onto the central star, and the disk radius increases.
As a result, the surface density decreases, and the gas cools further to maintain $Q \sim 1$. 
In a disk with $Q=1$, the Jeans mass is proportional to
\begin{equation}
M_{\rm Jeans} \propto \Sigma^3 \Omega^{-4}. 
\end{equation}
Therefore, as the surface density decreases, $M_{\rm Jeans}$ also decreases 
as $M_{\rm Jeans}\propto \Sigma^{3}$ and
the resolution requirement becomes more severe as the disk evolves. 
We can ignore the change in $\Omega$ due to  mass 
evolution of the central star because $M_{\rm star} \gg M_{\rm disk}$.

In an isolated disk with the local cooling law and 
without a temperature floor, the disk cools infinitely
to maintain $Q=1$, and the Jeans mass becomes infinitely 
small as the disk evolves (or as the surface density decreases). 
We must carefully monitor the Jeans mass during the simulations.

On the other hand, in a realistic system, there exists an appropriate range of temperature or of the Jeans mass,
depending on the system of interest.
We emphasize that the nature of the gravitational instability depends 
not only on the value of $Q$ but also on the most unstable wave-length of GI,
$\lambda_Q=\frac{2 c_s^2}{G \Sigma}$ ($\sim \lambda_{\rm Jeans}^2/H$), and 
different length scales give strikingly
different outcomes for the nonlinear growth phase. For example, the widths of 
spiral arms and masses of  
fragments differ for different length scales \citep[compare the spiral 
pattern in Fig. \ref{faceon_sigma} with Fig. 1 in ][]{2005MNRAS.364L..56R}.

We suggest that future self-gravitating disk simulations
should use a disk model that has a realistic Jeans length.
For example, the irradiated disk model \citep[e.g.,][]{2008ApJ...673.1138C} or a locally isothermal model
seems to be more suitable for numerical simulations
because we can limit the value of the 
most unstable wave-length to a realistic range.
To construct appropriate initial conditions, the profiles 
derived in \S 2 would be useful.

\section{Disk fragmentation and  evolution of clumps}
\label{fragment_section}
In this section, we investigate fragmentation of a disk and the evolution of clumps after their formation. Understanding the evolution of the clump is crucial
to consider the possible final outcomes 
(wide orbit planets, brown dwarfs or binary/multiple systems) that could result
from disk fragmentation.
\subsection{Simulation results}
\label{fragment_simulation_results}
\subsubsection{Parameter survey for disk fragmentation}
\label{parameter_range}

\begin{figure}
\includegraphics[width=80mm]{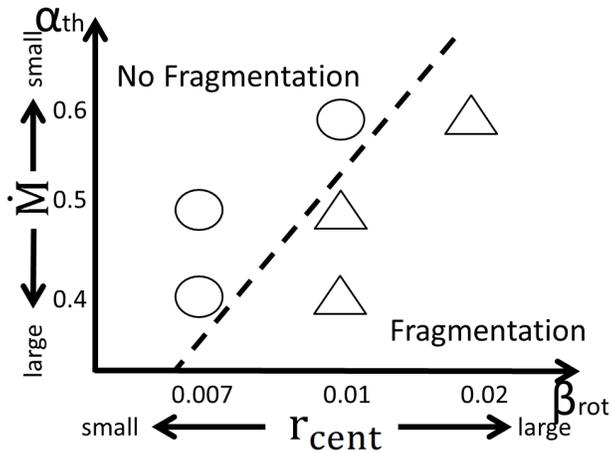}
\caption{
Classification of simulation results on the $\alpha_{\rm th}$-$\beta_{\rm rot}$ plane. Circles and 
triangles denote no-fragmentation and fragmentation models, respectively. Dashed line
shows the boundary between fragmentation and no fragmentation. 
}
\label{alpha_beta}
\end{figure}

As shown in Table 1, we calculated the evolution of the cloud core 
for five models.
In models 1, 3, and 5, fragmentation did not occur in the early phase of disk 
evolution. On the other hand, in models 2, 4, and TMI13, fragmentation did take place.
The computation results are summarized in Fig.~\ref{alpha_beta}. 
In that figure, circles (triangle) indicate models in which fragmentation 
did not (did) occur. The figure shows how fragmentation correlates with 
$\dot{M}$, the mass accretion rate onto the disk, and $r_{\rm cent}$, the centrifugal radius
of the fluid element. Qualitatively, larger mass accretion rates and larger centrifugal radii lead to fragmentation. 
The boundary that divides fragmentation and 
no-fragmentation models is consistent with that 
determined by simulations with the barotropic approximation \citep[see.,][]{2011MNRAS.416..591T}. 
This result suggests that, in the early evolution of a disk, 
fragmentation  rarely
depends on details of radiative transfer or thermodynamics of the gas. 
This is because disk fragmentation
in the early evolutionary phase is mainly determined by mass accretion rate onto the disk
and the angular momentum of mass accretion \citep{2003ApJ...595..913M,2006ApJ...645..381S,
2010ApJ...714L.133V,2010ApJ...719.1896V,
2010ApJ...718L..58I,2011MNRAS.416..591T}

\subsubsection{Difference of the conditions between fragmenting and non-fragmenting disk}
\label{diff_frag}
\begin{figure}
\includegraphics[width=50mm,angle=-90]{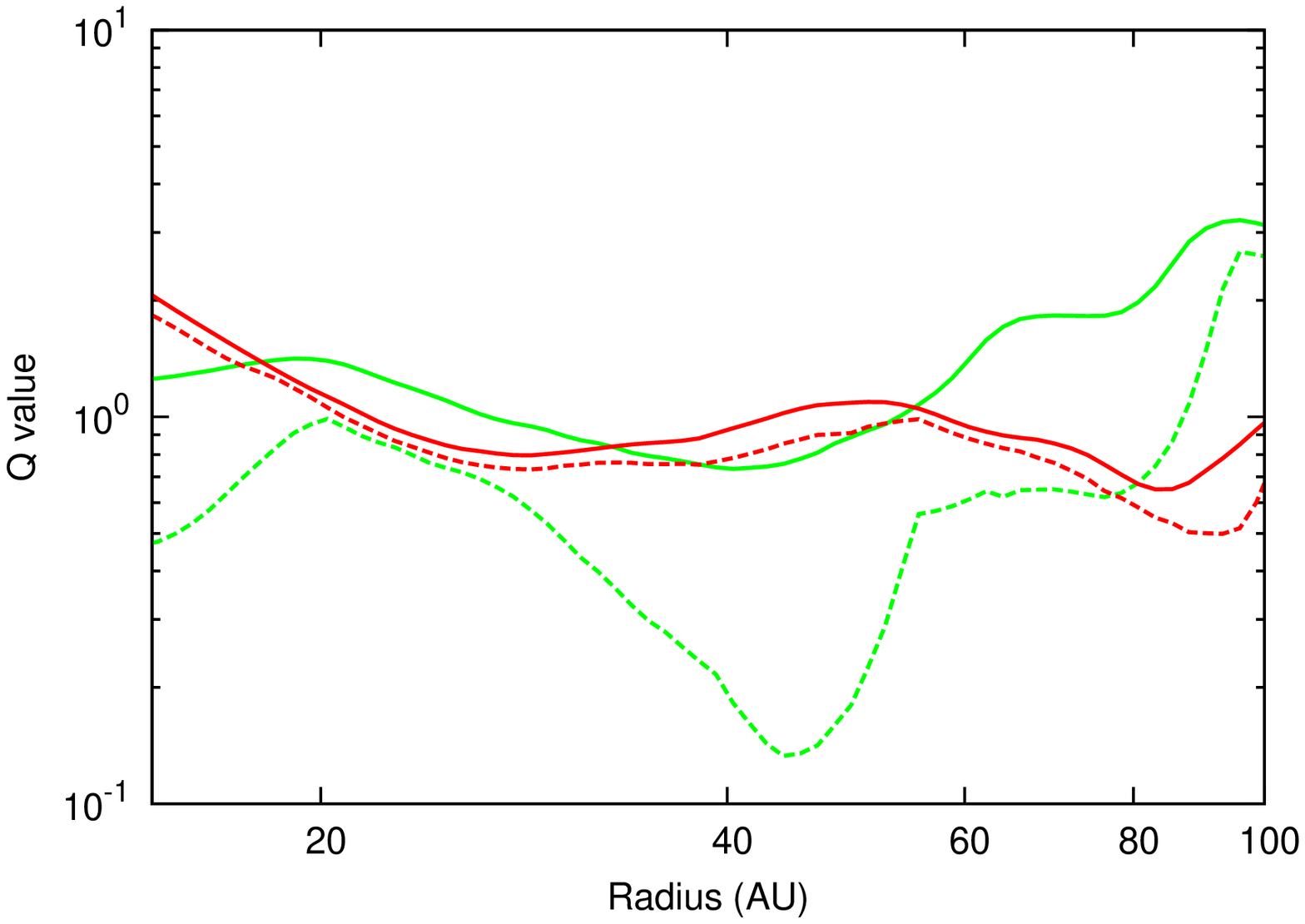} 
\includegraphics[width=50mm,angle=-90]{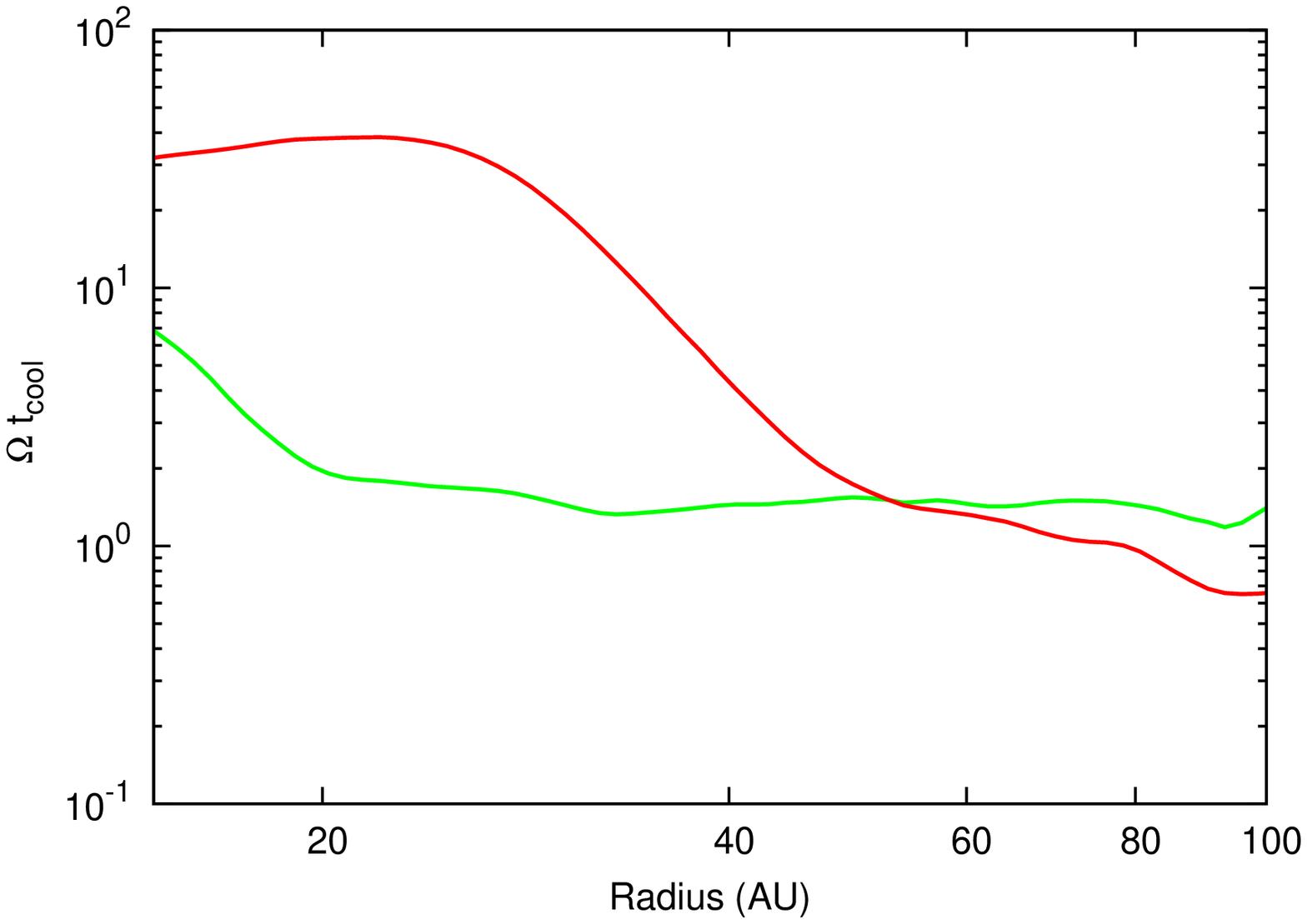} 
\caption{
Top panel shows the azimuthally averaged (solid) $Q$ value, $Q_{\rm ave}$ and the minimum $Q$ value, $Q_{\rm min}$ (dashed) of non-fragmenting (red; model 1) and
fragmenting disk (green; model 2).
Bottom panel shows the azimuthally averaged $\Omega t_{\rm cool}$ of non-fragmenting disk (red) and fragmenting disk (green).
The epoch of red lines is the same as in Fig.~\ref{faceon_sigma}{\it h}. 
The epoch of green lines is the same as in Fig.~\ref{faceon_sigma_frag}{\it a} and just prior to disk fragmentation.
}
\label{comp_profile}
\end{figure}

In this subsection, we investigate the difference of the conditions of fragmenting and non-fragmenting disk.
In Fig. \ref{comp_profile}, we show the $Q$ value and cooling time of model 1 (red lines; non-fragmenting disk) and 2 (green lines; fragmenting disk). 
The epoch of red lines is the same as in Fig.~\ref{faceon_sigma}{\it h}. 
The epoch of green lines is the same as in Fig.~\ref{faceon_sigma_frag}{\it a} and just prior to disk fragmentation.
The solid lines of Fig. \ref{comp_profile} show that the azimuthally averaged $Q$ value and the azimuthally averaged 
cooling time normalized by orbital period. They show both disks are gravitationally 
unstable ($Q_{\rm ave}\sim 1$) and their cooling time is sufficiently
small ($(\Omega t_{cool})_{\rm ave} \sim 1$). 
From the azimuthally averaged quantities, we can not 
find any significant difference between two disks.
But one fragments and the other does not.
What is the difference between them ?

The dashed lines in Fig. \ref{comp_profile} shows 
the minimum $Q$ value, $Q_{\rm min}$ at each radius. 
This traces the $Q$ value at the spiral arms.
We can see the clear difference between red and green dashed lines. 
The $Q_{\rm min}$ of fragmenting disk becomes 
significantly small ($Q_{\rm min}\lesssim 0.2$) in relatively 
large width ($\Delta r\sim 10$ AU).
On the other hand, the profile of $Q_{\rm min}$ in 
non-fragmenting disk does not show such a dip.
We monitored $Q_{\rm min}$ during entire period 
of evolution of non-fragmenting disk after protostar formation 
and found that $Q_{\rm min}$ never becomes $Q_{\rm min}< 0.2$.
Note also that the width of spiral arms of fragmenting disk is 
smaller than that of non-fragmenting disk 
(compare Fig.~\ref{faceon_sigma}{\it h} with 
Fig.~\ref{faceon_sigma_frag}{\it a}).

From these results, we conclude that the $Q$ value inside the 
spiral arms (not azimuthally averaged $Q$ value) and
the width of the spiral arms are important quantities for disk fragmentation. 
The importance of width of spiral arms is also pointed out by \citet{2012MNRAS.423.1896R}.

\subsubsection{Fragmentation of disks and evolution of the clumps}
\label{sec4_clump}
In this subsection, we consider the evolution of the fragments 
(or clumps) formed by disk fragmentation.
We have already investigated the evolution of clumps in \citet{2013MNRAS.436.1667T}. 
The parameters for the cloud core investigated in \citet{2013MNRAS.436.1667T} 
were $\alpha_{\rm th}=0.3844$ and $\beta_{\rm rot}=0.01$ (model TMI13 in Table 1).
In model TMI13, low $\alpha_{\rm th}$ and $\beta_{\rm rot}$ were adopted. This means that
the disk formed in model TMI13 is compact.
In that case, fragmentation occurred in the region $T\gtrsim 20$ K. 
Thus, the entropy of the centers of clumps
is greater than that of the typical first-core \citep[see Fig. 4 in][]{2013MNRAS.436.1667T}.

When the initial cloud core has a high value of $\beta_{\rm rot}$, it is expected that 
the disk will be more outspread, and clumps will form in low-temperature regions. Therefore, the 
entropy of the clumps would become small. The central entropy of clumps is important when 
we consider the initial mass and evolution of clumps.
Smaller entropy leads to smaller clump mass (see eq. (\ref{critical_mass})) and a smaller clump radius.
Thus, investigating clump evolution for larger value of $\alpha_{\rm th}$ and $\beta_{\rm rot}$ would be insightful to understand the 
initial properties and the evolutionary process of clumps in more detail.
To do so, we investigate clump evolution in model 2.  
The cloud core of model 2 has $\alpha_{\rm th}$ and $\beta_{\rm rot}$ values greater than those in model TMI13.

\begin{figure}
\includegraphics[width=90mm]{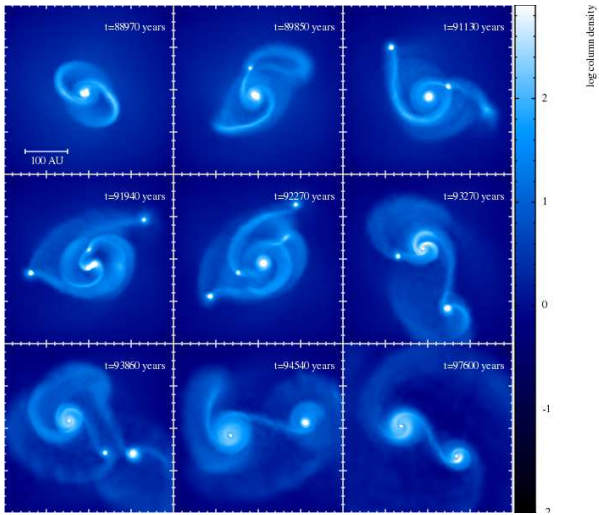}

\caption{
Time sequence for  surface density at the cloud center, viewed face-on for model 2.
Elapsed time after the cloud core begins to collapse is shown in each panel. 
}
\label{faceon_sigma_frag}
\end{figure}

Figure \ref{faceon_sigma_frag} shows the evolution of surface density at the center of the cloud core for model 2.
Analogous to Fig.~\ref{faceon_sigma}, Fig.~\ref{faceon_sigma_frag}{\it a} shows bimodal structure
(the first-core and disk around it) in the early phase of disk evolution. 
The first clump formed in Fig.~\ref{faceon_sigma_frag}{\it b}. 
We define the epoch of clump formation as the time when its central density
exceeds $10^{-11}~{\rm g\,cm^{-3}}$.
Within $8000$ years after the first clump formation, six more clumps formed.

\begin{figure}
\includegraphics[width=90mm]{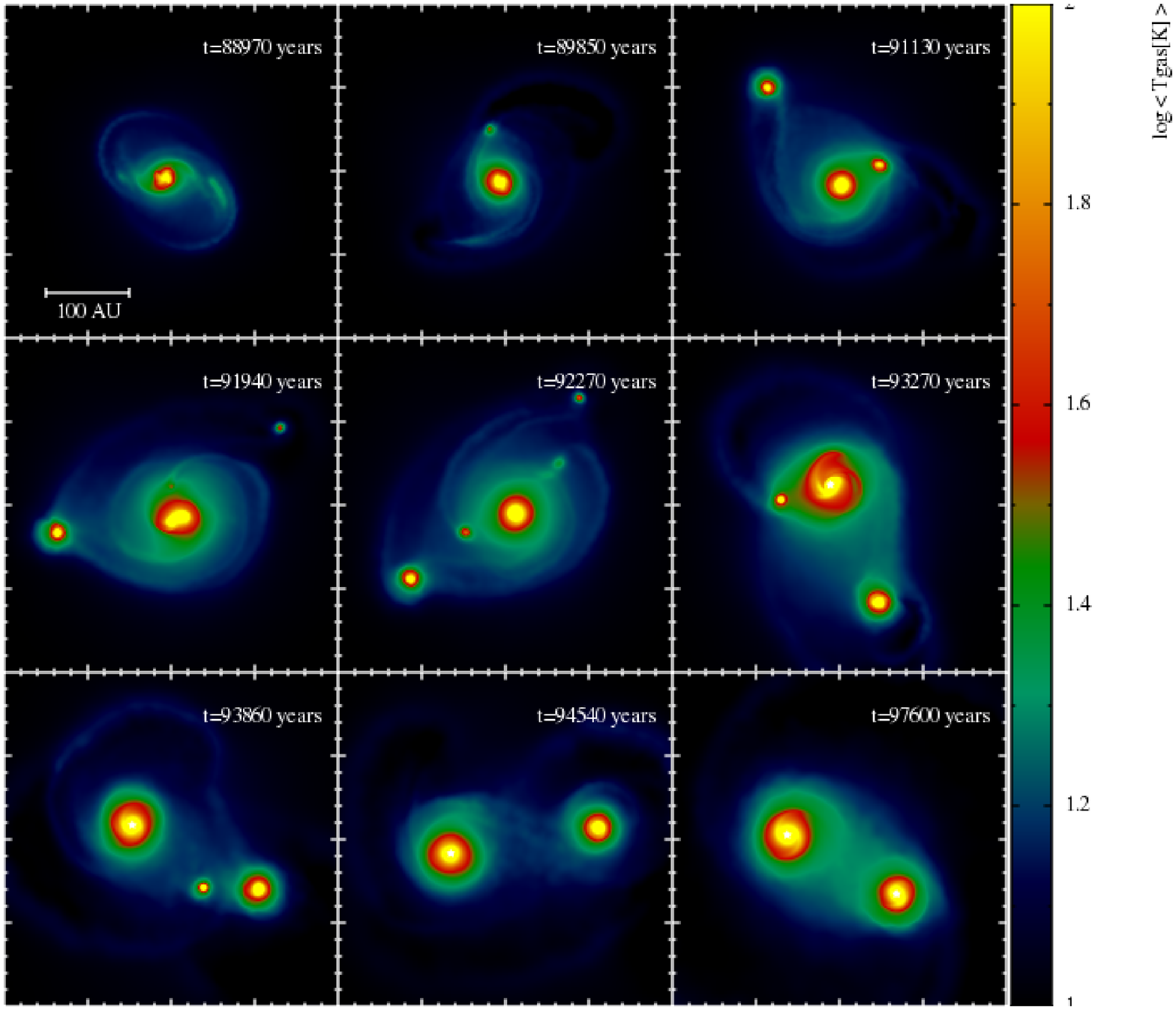}

\caption{
Time sequence for the density-weighted gas temperature at the cloud center, viewed face-on for model 2.
Elapsed time after the cloud core begins to collapse is shown in each panel. 
}
\label{faceon_Tgas_frag}
\end{figure}

Figure \ref{faceon_Tgas_frag} shows the temperature evolution at the center of the cloud core of model 2.
As expected, disk fragmentation and clump formation occur in the very low-temperature region of
 $T\sim 10$ K. After disk fragmentation, the central temperature 
of the clump quickly increases because of compressional heating.

\begin{figure}
\includegraphics[width=70mm,angle=-90]{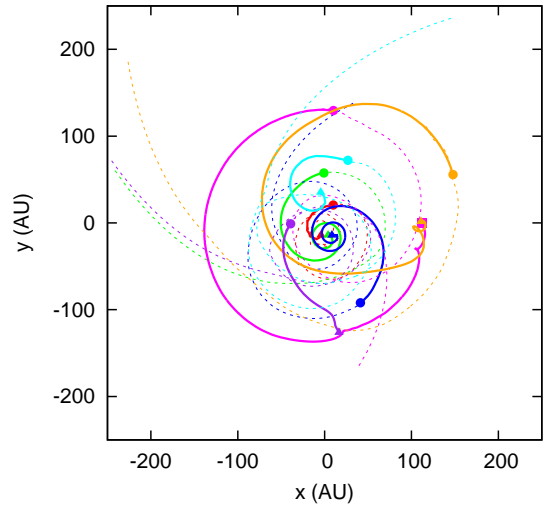}

\caption{
Orbits of clumps (solid lines).
Dashed lines represent orbits of the fluid element of the clump center before clump 
formation ($\rho_c<10^{-11} \cm$).
Symbols mark positions where clumps form (circles), the clump central
density begins to decrease or the clump begins to be destroyed (triangles), and second collapse begins in the clump or the 
sink particles are inserted (square).
}
\label{x_y_clump}
\end{figure}

Figure \ref{x_y_clump} shows the orbits of clumps after clump formation (solid lines) and
the fluid elements at the centers of the clumps prior to clump formation
(dashed lines). 
Four clumps, shown with red, green, blue, and cyan lines,
finally accreted onto the central first-core or protostar. 
On the other hand, the remaining three clumps, shown with magenta, purple, 
and orange lines merged into one clump and became a secondary 
protostar (see, Fig.~\ref{faceon_sigma_frag}{\it i}).

\begin{figure}
\includegraphics[width=60mm,angle=-90]{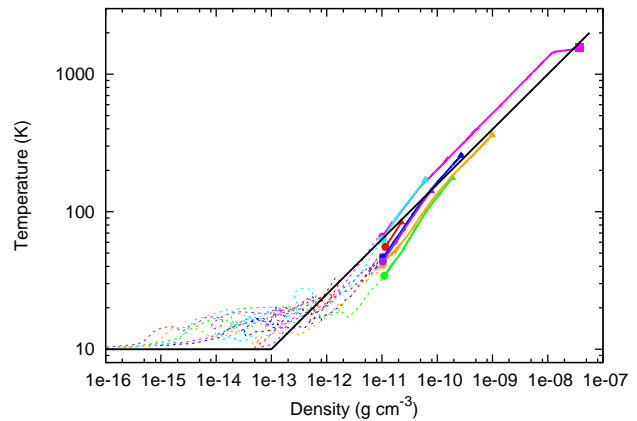}

\caption{
Temperature at the centers of clumps (solid lines) as a function of 
density. 
Dashed lines represent the temperature evolution of fluid elements at clump centers before clump 
formation ($\rho_c<10^{-11} \cm$).
Symbols and colors are the same as in Fig.~\ref{x_y_clump}.
A typical evolution of the barotropic approximation is shown by black solid line.
}
\label{rho_T_clump}
\end{figure}

Figure \ref{rho_T_clump} shows the temperature evolution of the centers of the clumps
as a function of density (solid lines). 
The figure also shows the temperature evolution of fluid elements at the centers of the clumps prior to clump formation
(dashed lines).
In this figure, a typical temperature evolution with the barotropic equation of state, which
is designed to mimic the temperature evolution of the center of the 
first-core \citep[][]{2000ApJ...531..350M}, is also shown for comparison.

Until the density reaches $\sim 10^{-13} \cm$, the temperature is 
almost isothermal because there is no significant heating
by spiral arms. On the other hand,  above $\sim 10^{-13} \cm$, the gas evolution
becomes adiabatic.
The interesting finding is that the thermal evolution of the central temperature 
is almost consistent with the evolution in the barotropic approximation  (black solid line).
Because clumps form in the disk, we might expect that fluid elements at the centers of clumps
have a good chance of decreasing their entropy by radiative cooling
compared with that of the first-core, which is surrounded by a spherically symmetric massive envelope. 
However, our results show that centers of clumps do not efficiently cool
before they become optically thick.  As a result, the evolution of fluid elements
follows the line of the barotropic equation of state.

\begin{figure}
\includegraphics[width=60mm,angle=-90]{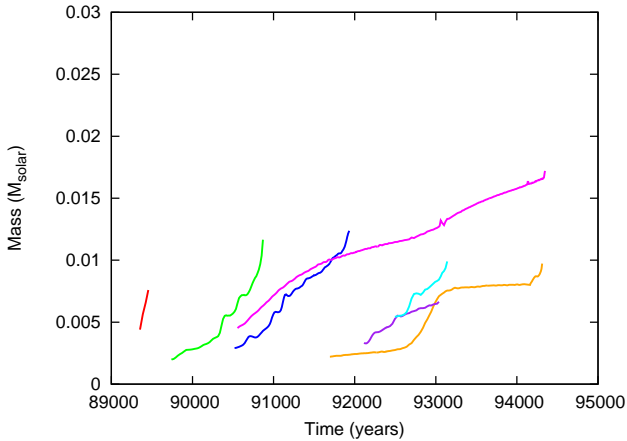}

\caption{
Masses of clumps as  functions of elapsed time after the initial molecular cloud core begins to
collapse.
Colors have the same meaning as in Figs.~\ref{x_y_clump} and \ref{rho_T_clump}.
}
\label{time_mass_clump}
\end{figure}

Figure \ref{time_mass_clump} shows how the mass of 
clumps evolves over time. Here, we define the mass of a clump, 
$M_c$, so that it satisfies
\begin{eqnarray}
\label{def_mass}
3\int^{M_c}_{0} \frac{p}{\rho}dM_r=\int^{M_c}_{0}\frac{GM_r}{r}dM_r,
\end{eqnarray}
where $p,~\rho$ and $M_r$ represent the pressure, density, and cumulative mass at $r$, respectively, and 
$r$ is the radius from the clump center.
Note that (\ref{def_mass}) is identical to the Virial theorem 
when the surface pressure is negligible, and the clump is
in hydrostatic equilibrium.
Just after clump formation, the mass of each clump is 
a few Jupiter mass. This mass is slightly smaller than that of the typical clump formed in
\citet{2013MNRAS.436.1667T}  because the central entropy of the clump formed in model 2 is smaller 
than that in \citet{2013MNRAS.436.1667T}. 
The mass of the clump indicated by orange in Fig. \ref{time_mass_clump}
quickly increased at $t\sim 93,000$ years. This is because the clump encountered
a disk spiral arm and gained  mass from it.
The clumps indicated by purple and orange merged 
into the clump indicated by magenta. Even with this coalescence,
the mass of the clump indicated by magenta did not increase significantly. This is because the 
clumps indicated by purple and cyan were
tidally destroyed and formed circum-clump 
disk around the clump indicated by magenta. Therefore, the mass of the destroyed 
clumps did not directly accrete onto the clump indicated by magenta.

In the clump indicated by magenta, the second collapse occurred about 3000 years after its formation. 
This timescale is consistent with our previous results \citep{2013MNRAS.436.1667T}. 
Note that the second collapse occurs at a slightly smaller mass ($\sim 0.015 M_\odot$) compared to that in
our previous results. This is because the
central entropy of the clump is smaller than that of the clumps in \citet{2013MNRAS.436.1667T}. 
The smaller central entropy 
induces the second collapse at a smaller clump mass \citep[see (3) in][]{2013MNRAS.436.1667T}.


\subsection{Discussion}
\label{fragment_simulation_discussion}

\subsubsection{Minimum central entropy of clumps}
\label{minimum_entropy}

An important finding in \S \ref{sec4_clump} is that the evolutionary paths of 
the central temperatures of clumps on the $\rho-T$ plane are close to that of the first-core even though the 
clumps form in the disk.
This implies that the clump evolution after the fragmentation can be approximately regarded as 
spherically-symmetric.
Therefore, in the same manner as in the case of the first-core, 
we can describe the evolution of the central entropy of the clump.

Following \citet{1999ApJ...510..822M} and \citet{2005ApJ...626..627O},
the minimum entropy is calculated as follows.
We assume the quasi-adiabatic evolution begins when the clump becomes opaque.
If we assume the clump radius is the Jeans length $\lambda_{\rm Jeans}=\sqrt{\pi c_s^2/(G \rho)}$,
the clump becomes opaque when
\begin{equation}
\label{omukai2}
\tau_J=\kappa \rho \lambda_{\rm Jeans} \sim1,
\end{equation}
is satisfied.

Because we are interested in the minimum central entropy, we neglect
heating by the irradiation.
In such a case, the clump evolves according to the energy balance between
optically thin radiative cooling and compressional heating until it becomes opaque.
The energy balance can be written as
\begin{equation}
\label{omukai1}
\kappa \sigma T^4=\frac{c_s^2}{t_{ff}},
\end{equation}
where $t_{ff}=\sqrt{3 \pi/(32 G \rho)}$ is the free-fall timescale.
We assume the opacity is given as
\begin{equation}
\label{opacity}
\kappa = \kappa_0 \left(\frac{T}{10 {\rm K}} \right)^2 ~ {\rm cm^2 ~g^{-1}}.
\end{equation}
Using equations (\ref{omukai2}), (\ref{omukai1}), and (\ref{opacity}), the density and temperature where
the adiabatic evolution begins, $\rho_{\rm ad}$ and $T_{\rm ad}$ are given as
\begin{eqnarray}
\rho_{\rm ad} &=& 7.6 \times 10^{-14} \left(\frac{\kappa_0}{0.01~ {\rm cm^2 ~g^{-1}}}\right)^{-2/3} \cm, \\
T_{\rm ad} &=& 15  \left( \frac{\kappa_0}{0.01~ {\rm cm^2 ~g^{-1}}} \right)^{-4/15} {\rm K}.
\end{eqnarray}
The minimum value of entropy at the center of the clump is given by these values.

Note that the minimum entropy does not depend on $\kappa_0$ when the ratio 
of heat capacities $\gamma$ is equal to $7/5$ because
$ T_{\rm ad}/\rho_{\rm ad}^{\gamma-1} \propto \kappa_0^{3/2(\gamma - 7/5)}$ \citep{2005ApJ...626..627O}.
Therefore, the evolution of the central region of the clump follows the same track in the $\rho-T$ plane irrespective 
of different $\kappa$. The property is the same as in the evolution of the protostars with various
metallicities shown by \citet{2005ApJ...626..627O}.

\subsubsection{Minimum initial mass of the clump}
\label{minimum_entropy2}
{\rm
Based on the discussions in \S \ref{minimum_entropy}}, we can evaluate the minimum initial 
mass of a clump by assuming that the clump can be described by a
polytropic sphere \citep[as we have shown in ][ the clump 
can be well described with polytropic sphere with $n\sim 3- 4$]{2013MNRAS.436.1667T}. 
the mass of the clump with polytropic index $n$ 
is calculated by
\begin{equation}
\begin{split}
\label{critical_mass}
M_{\rm min}=&(n+1)^{3/2} \left[ \frac{k_{\rm B}^3}{4\pi G \mu^3m_{\rm H}^3}\frac{T_c^3}{\rho_c}\right]^{\frac{1}{2}} \left[- \xi^2 \frac{d \theta}{d \xi} \right]_{\xi=\xi_n} \\
=&\left\{\
     \begin{array}{c}
      5.4 \times 10^{-3}    
 ~(n=4) \\
      4.3 \times 10^{-3}    
 ~(n=3)
     \end{array}
     \right\} \\
 & \times  \left(\frac{T_c}{15 {\rm K}}\right)^{\frac{1}{4}}\left(\frac{\rho_{{\rm ad}}}{7.6 \times 10^{-14}
       \cm}\right)^{-\frac{1}{2}}\left(\frac{T_{{\rm ad}}}{15 {\rm K}}\right)^{\frac{5}{4}}
 M_{\odot},
\end{split}
\end{equation}
where $T_c$, $k_{\rm B}$, and $\mu$ are the central temperature, Boltzmann constant, and  mean molecular weight, respectively.
In the estimate, the heat capacity at constant volume $c_v$ and the ratio 
of heat capacities, $\gamma$ ($=7/5$)
are assumed to be constant for simplicity. 
{\rm The estimated initial mass is a few Jupiter mass}, which is consistent with 
the initial mass of the clumps formed in our simulation (see, Fig. \ref{time_mass_clump}).

\subsubsection{Comparison with previous work}
There are few studies about the evolution of the clumps with 
realistic accretion onto them and sufficient numerical resolution to resolve the central structure of clump
\citep{2009MNRAS.400.1563S,2013MNRAS.436.1667T}.
In this subsection, we compare our results with those in \citet{2009MNRAS.400.1563S}.

\citet{2009MNRAS.400.1563S} investigated the evolution of the clumps with three-dimensional
simulation starting from a massive isolated disk.
In the disk, fragmentation immediately occurs and the clumps form.
They show that the clump undergoes the second collapse when its mass reaches
$\sim 10 M_{\rm Jupiter}$ and the timescale for the second collapse is several thousand years.
They also points out that thermal evolution of the clump is consistent 
with the evolution of first-core \citep{2000ApJ...531..350M}.
The evolution process of the clumps formed in our simulation is consistent with that found in
their simulations.

However, there are interesting differences between their results and ours. In their simulations, most of the clumps
remained in the disk  without falling onto the central protostar. On the other hand, in our simulation, a large fraction 
 of clumps fell onto the central star and disappeared. 
This difference may come from the fact that the initial disk of \citet{2009MNRAS.400.1563S} 
is very massive ($0.7 M_\odot$) and unstable ($Q\sim0.9$). 
In such a massive disk, fragmentation immediately occurs and many clumps simultaneously form.
Thus, many clumps reside in the disk at the same epoch and interact each other.
On the other hand, in our simulation, only two or three clumps simultaneously exist in the
disk at the same epoch because the disk is not so massive. The difference in the orbital evolution of the clump may 
come from the difference of the disk.
Orbital evolutions of clumps depend more strongly on the condition of 
disk and further studies are needed on this issue.

\subsubsection{Possible evolutionary path to realize the small mass clump}

Our simulation results have shown that the central entropy of a clump
cannot become significantly smaller than that of the first core.
However, there is a possible path that can lead to a clump with an central entropy smaller than that of the first core.
The above discussion relies on  the clump evolution  {\it after}  disk fragmentation.
Thus, the value of $Q$ in the disk has already become $Q\sim 1$.
Once gravitational instabilities turns on, the surface 
density decreases by mass and angular momentum
transfer by the spiral arms, or disk fragmentation occurs.
As shown in \S \ref{sec4_clump}, the rapid evolution of the clump after fragmentation 
prevents temperature evolution  which leads to smaller central entropy than that of the first core.

However, when $Q$ is greater than unity, the disk surface density can monotonically increase by in-fall from the envelope
without mass and angular momentum re-distribution or fragmentation by GI in the disk.
The maximum mid-plane density of a gravitationally stable Keplerian disk ($Q\gtrsim1$) 
can be calculated from the condition $Q=1$ as
\begin{equation}
\rho_{\rm max}=\frac{c_s \Omega }{\pi G H } = 1.88 \times 10^{-10}  
\left( \frac{M_{\rm star}}{M_\odot} \right) \left( \frac{r}{10 {\rm AU}} \right)^{-3} \cm,
\end{equation}
where, $M_{\rm star}$ is the mass of the central star.
The maximum mid-plane density is solely determined by the angular velocity. 
 
This indicates that a high mid-plane density can be achieved in the inner region of the disk.
If the disk with such a high mid-plane density at the inner region and low temperature somehow fragments,
a clump with small central entropy could can be created.
For example, if a disk with a temperature of 50 K at 10 AU around $1 M_{\odot}$ fragments,
adiabatic contraction of the clump begins from $\rho \sim 10^{-10} \cm$ and $~T=50 $ K. In such a case, 
the initial mass of clump is $\sim 0.7~ M_{\rm Jupiter}$. 
However, the realization of such a disk seems to 
require the inclusion of magnetic field 
\citep{2010ApJ...718L..58I,2011ApJ...729...42M}, which is beyond the scope of this paper.

\section{Summary and future study}
\label{summary_future}
\subsection{Summary}
In this paper, we investigate the structure of self-gravitating disks, their fragmentation and the 
evolution of the fragments using an analytic approach and  three-dimensional radiation hydrodynamics
simulations.
First, we analytically 
derive the quasi-steady structures of self-gravitating disks with various energy equations.
We show that local cooling law, which has been widely used in previous studies
\citep[e.g.,][]{2005MNRAS.364L..56R,2005MNRAS.358.1489L,2011MNRAS.416L..65P,2012MNRAS.427.2022M}, 
describe a globally isothermal disk ($T \propto r^0$) in the quasi-steady state. 
In addition, we point out that Jeans mass of the disk with local cooling law 
and the fiducial disk model ($M_{star}=1,~M_{disk}=0.1,~r_{in}=0.25$ and $r_{out}=25$) becomes very small
and resolution requirement for the simulation is exceedingly severe. We show that at least 
$\sim 10^7$ particles are required to resolve Jeans mass of the disk.
We also point out that such a small Jeans mass ($\sim 10^{-5} M_\odot$, if we regard the 
mass of the central star as $1~M_\odot$) are not attainable in a realistic disk around low mass star since it 
requires unrealistically low temperature.

We also investigated the quasi-steady structure of the disk in which radiative cooling 
locally balances viscous heating and showed that it has very steep 
radial profiles (e.g., $T\propto r^{-3}$ for Keplerian disk) in the quasi-steady state.
To investigate whether such a steep profile can be realized (in other words, whether the local balance between 
radiative cooling and viscous heating can be realized), 
we conducted three-dimensional radiation hydrodynamics simulations. We found that the disk does not have the
radial profile expected from the local energy balance and that the disk temperature
is non-locally determined by radiative transfer.
The radial profile of the disk temperature was
$T \propto  r^{-1.1}$ and this scaling can be derived analytically by applying diffusion approximation.
Because the temperature of the disk is determined by radial radiative transfer within the disk, 
radial radiation transport is crucial for outer regions of the disk.
Thus, we conclude that the description only with local radiative cooling is not viable in massive disks 
around low mass stars

The disk formed in our simulation satisfies the 
fragmentation criterion based on disk cooling time ($Q \sim 1$ and $\Omega t_{\rm cool} \sim 1$).
However, the fragmentation is not observed in the disk.
Therefore, the fragmentation criterion is not sufficient condition for fragmentation.
Further studies for more accurate criterion is needed.

Mass accretion from the envelope is another process that possibly drives disk fragmentation.
By a parameter survey with radiation hydrodynamics simulations starting from the molecular cloud core,
we determined the parameter range of the host cloud core needed for the gravitational fragmentation. The parameter range 
is in agreement with that obtained in our previous study based on a simplified equation of state \citep{2011MNRAS.416..591T}. 
Therefore, we conclude detailed treatment of radiative transfer 
is not crucial for disk fragmentation driven by mass accretion from the envelope.

We also investigated the internal evolution of fragments (clumps) formed in extended 
cold disks ($T\sim 10$ K in the outer region). 
Even in such a disk, the central temperature of a clump does 
not sufficiently cool to  have smaller central entropy than that of the first core.
Therefore,  there is  a minimum value of the central entropy.
Using this value for the central entropy, we derived the minimum initial mass 
of the clumps to be about a few Jupiter mass.
This is consistent with the initial mass formed in our simulations.

\subsection{Future study}
In our simulations, the flux limited diffusion (FLD) approximation was adopted.
It is well known that the FLD approximation does not  behave well in optically thin region.
Although the disk formed in our simulation can be
regarded as optically thick in almost all regions,

it is possible that the temperature profile
changes with more realistic radiative transfer methods
because the simulation box includes the optically thin region and the radiation flows along  a roundabout path in the
FLD approximation. 
Although the difference is expected to be about a factor of a few
\cite[see comparison tests of radiative transfer schemes; e.g.,][]{2004A&A...417..793P,2013A&A...555A...7K} 
, it is important to perform simulations with more sophisticated radiative transfer method.
We plan to perform simulations with more sophisticated radiative transfer schemes 
to confirm our results for the disk structure of a massive disk around a low mass star.

Another important issue, which is not discussed in this paper, is the effects of magnetic fields.
Magnetic fields play important roles in the formation and early evolution of circumstellar disks due to their efficient
transfer of angular momentum and formation of outflows
\citep[see,][ and references therein]{2012PTEP.2012aA307I}.
In particular,  magnetic fields would change the parameter range of fragmentation that is derived in \S \ref{parameter_range}.
It would also be important for orbital and internal evolutions of clumps. The disk is well coupled to 
magnetic fields because magnetic diffusion is not significant in the range of disk densities. On the other hand,
the magnetic diffusion is effective inside clumps because of their high density. 
Thus, the magnetic field and clumps would be decoupled.
In our future study, we will investigate the effects of magnetic fields 
using numerical methods for smoothed particle magneto-hydrodynamics (SPMHD) as described in
\citet[][]{2011MNRAS.418.1668I} and \citet{2013MNRAS.434.2593T}.

\section *{Acknowledgments}
We thank  K. Iwasaki, T. Matsumoto, H. Kobayashi, M. Ogihara, K. Tanaka and T. Muto for their fruitful discussions.
We also thank K. Tomida and Y. Hori to provide their EOS table to us.
We also thank the referee, Dr. D. Stamatellos for insightful comments.
The snapshots were produced by SPLASH \citep{2007PASA...24..159P}.
The computations were performed on a parallel computer, XC30 system at CfCA of
NAOJ and SR16000 at YITP in Kyoto University.
Y.T. is financially supported by Research Fellowships of JSPS for Young Scientists.

\bibliography{article}

\end{document}